\pdfoutput=1
\documentclass[12pt,a4paper]{article}
\usepackage[colorinlistoftodos]{todonotes}

\usepackage{ifthen} % for conditional statements
\newboolean{pdflatex}
\setboolean{pdflatex}{true} % False for eps figures 

\newboolean{articletitles}
\setboolean{articletitles}{true} % False removes titles in references

\newboolean{uprightparticles}
\setboolean{uprightparticles}{false} %True for upright particle symbols

\newboolean{inbibliography}
\setboolean{inbibliography}{false} %True once you enter the bibliography

\def\paperauthors{LHCb collaboration} % Leave as is for PAPER and CONF
\def\paperasciititle{Measurement of the branching fractions of the decays D+->KKK, D+->pipiK and D+_s->piKK} % Set ASCII title here
\def\papertitle{Measurement of the branching fractions of the decays \dkkk,  \dpipik and \dspikk} % Latex formatted title
\def\paperkeywords{{High Energy Physics}, {LHCb}} % Comma separated list
\def\papercopyright{\the\year\ CERN for the benefit of the LHCb collaboration} % new since 9/Apr/2018
\def\paperlicence{CC-BY-4.0 licence}
\def\paperlicenceurl{https://creativecommons.org/licenses/by/4.0/}

\usepackage[top=1in, bottom=1.25in, left=1in, right=1in]{geometry}

\usepackage{microtype}
\usepackage{lineno}  % for line numbering during review
\usepackage{xspace} % To avoid problems with missing or double spaces after
\usepackage{caption} %these three command get the figure and table captions automatically small

\usepackage{graphicx}  % to include figures (can also use other packages)
\usepackage{color}
\usepackage{colortbl}
\graphicspath{{./figs/}} % Make Latex search fig subdir for figures

\usepackage{amsmath} % Adds a large collection of math symbols
\usepackage{amssymb}
\usepackage{amsfonts}
\usepackage{upgreek} % Adds in support for greek letters in roman typeset

\newcommand*\patchAmsMathEnvironmentForLineno[1]{%
\expandafter\let\csname old#1\expandafter\endcsname\csname #1\endcsname
\expandafter\let\csname oldend#1\expandafter\endcsname\csname
end#1\endcsname
 \renewenvironment{#1}%
   {\linenomath\csname old#1\endcsname}%
   {\csname oldend#1\endcsname\endlinenomath}%
}
\newcommand*\patchBothAmsMathEnvironmentsForLineno[1]{%
  \patchAmsMathEnvironmentForLineno{#1}%
  \patchAmsMathEnvironmentForLineno{#1*}%
}
\AtBeginDocument{%
\patchBothAmsMathEnvironmentsForLineno{equation}%
\patchBothAmsMathEnvironmentsForLineno{align}%
\patchBothAmsMathEnvironmentsForLineno{flalign}%
\patchBothAmsMathEnvironmentsForLineno{alignat}%
\patchBothAmsMathEnvironmentsForLineno{gather}%
\patchBothAmsMathEnvironmentsForLineno{multline}%
\patchBothAmsMathEnvironmentsForLineno{eqnarray}%
}

\usepackage{hyperxmp}

\usepackage[pdftex,
            pdfauthor={\paperauthors},
            pdftitle={\paperasciititle},
            pdfkeywords={\paperkeywords},
            pdfcopyright={Copyright (C) \papercopyright},
            pdflicenseurl={\paperlicenceurl}]{hyperref}

\usepackage[all]{hypcap} % Internal hyperlinks to floats.

\usepackage{xspace} 
\usepackage{upgreek}

\def\lhcb {\mbox{LHCb}\xspace}

\def\MagUp {\mbox{\em Mag\kern -0.05em Up}\xspace}
\def\MagDown {\mbox{\em MagDown}\xspace}

\ifthenelse{\boolean{uprightparticles}}%
{

 \def\Ppi         {\ensuremath{\uppi}\xspace}

 \def\PDelta      {\ensuremath{\Delta}\xspace}                 
 \def\PXi      {\ensuremath{\Xi}\xspace}                 
 \def\PLambda      {\ensuremath{\Lambda}\xspace}                 
 \def\PSigma      {\ensuremath{\Sigma}\xspace}                 
 \def\POmega      {\ensuremath{\Omega}\xspace}                 
 \def\PUpsilon      {\ensuremath{\Upsilon}\xspace}                 
 
 \def\PB      {\ensuremath{\mathrm{B}}\xspace}                 
                  
 \def\PD      {\ensuremath{\mathrm{D}}\xspace}

 \def\PK      {\ensuremath{\mathrm{K}}\xspace}

 \def\Pb      {\ensuremath{\mathrm{b}}\xspace}                 
 \def\Pc      {\ensuremath{\mathrm{c}}\xspace}

 \def\Pi      {\ensuremath{\mathrm{i}}\xspace}

 \def\Ps      {\ensuremath{\mathrm{s}}\xspace}

}
{

 \def\Ppi         {\ensuremath{\pi}\xspace}

 \mathchardef\PDelta="7101
 \mathchardef\PXi="7104
 \mathchardef\PLambda="7103
 \mathchardef\PSigma="7106
 \mathchardef\POmega="710A
 \mathchardef\PUpsilon="7107
                  
 \def\PB      {\ensuremath{B}\xspace}                 
                  
 \def\PD      {\ensuremath{D}\xspace}

 \def\PK      {\ensuremath{K}\xspace}

 \def\Pb      {\ensuremath{b}\xspace}                 
 \def\Pc      {\ensuremath{c}\xspace}

 \def\Pi      {\ensuremath{i}\xspace}

 \def\Ps      {\ensuremath{s}\xspace}

}

\makeatletter
\ifcase \@ptsize \relax% 10pt
  \newcommand{\miniscule}{\@setfontsize\miniscule{4}{5}}% \tiny: 5/6
\or% 11pt
  \newcommand{\miniscule}{\@setfontsize\miniscule{5}{6}}% \tiny: 6/7
\or% 12pt
  \newcommand{\miniscule}{\@setfontsize\miniscule{5}{6}}% \tiny: 6/7
\fi
\makeatother

\DeclareRobustCommand{\optbar}[1]{\shortstack{{\miniscule (\rule[.5ex]{1.25em}{.18mm})}
  \\ [-.7ex] $#1$}}

\def\squark    {{\ensuremath{\Ps}}\xspace}

\def\cquark    {{\ensuremath{\Pc}}\xspace}

\def\bquark    {{\ensuremath{\Pb}}\xspace}

\def\pion   {{\ensuremath{\Ppi}}\xspace}

\def\pip    {{\ensuremath{\pion^+}}\xspace}
\def\pim    {{\ensuremath{\pion^-}}\xspace}

\def\kaon    {{\ensuremath{\PK}}\xspace}
  \def\Kbar    {{\kern 0.2em\overline{\kern -0.2em \PK}{}}\xspace}

\def\KorKbar    {\kern 0.18em\optbar{\kern -0.18em K}{}\xspace}
\def\Kz      {{\ensuremath{\kaon^0}}\xspace}
\def\Kzb     {{\ensuremath{\Kbar{}^0}}\xspace}
\def\Kp      {{\ensuremath{\kaon^+}}\xspace}
\def\Km      {{\ensuremath{\kaon^-}}\xspace}

\def\KS      {{\ensuremath{\kaon^0_{\mathrm{ \scriptscriptstyle S}}}}\xspace}

  \def\Dbar    {{\kern 0.2em\overline{\kern -0.2em \PD}{}}\xspace}
\def\D       {{\ensuremath{\PD}}\xspace}

\def\DorDbar    {\kern 0.18em\optbar{\kern -0.18em D}{}\xspace}
\def\Dz      {{\ensuremath{\D^0}}\xspace}

\def\Dp      {{\ensuremath{\D^+}}\xspace}

\def\Dps      {\ensuremath{\D^+_{(\squark)}}\xspace}
\def\Dms      {\ensuremath{\D^-_{(\squark)}}\xspace}
\def\Dsp     {{\ensuremath{\D^+_\squark}}\xspace}

\def\B       {{\ensuremath{\PB}}\xspace}
\def\Bbar    {{\ensuremath{\kern 0.18em\overline{\kern -0.18em \PB}{}}}\xspace}

\def\BorBbar    {\kern 0.18em\optbar{\kern -0.18em B}{}\xspace}

  \def\Y#1S{\ensuremath{\PUpsilon{(#1S)}}\xspace}% no space before {...}!

\def\Lz          {{\ensuremath{\PLambda}}\xspace}
\def\Lbar        {{\ensuremath{\kern 0.1em\overline{\kern -0.1em\PLambda}}}\xspace}
\def\LorLbar    {\kern 0.18em\optbar{\kern -0.18em \PLambda}{}\xspace}

\def\Lc      {{\ensuremath{\Lz^+_\cquark}}\xspace}

\def\BF         {{\ensuremath{\mathcal{B}}}\xspace}

\newcommand{\decay}[2]{\ensuremath{#1\!\to #2}\xspace}         % {\Pa}{\Pb \Pc}

\def\to                 {\ensuremath{\rightarrow}\xspace}

\def\AT#1     {\ensuremath{A_{\mathrm{T}}^{#1}}\xspace}           % 2

\def\C#1      {\ensuremath{\mathcal{C}_{#1}}\xspace}                       % 9
\def\Cp#1     {\ensuremath{\mathcal{C}_{#1}^{'}}\xspace}                    % 7
\def\Ceff#1   {\ensuremath{\mathcal{C}_{#1}^{\mathrm{(eff)}}}\xspace}        % 9  
\def\Cpeff#1  {\ensuremath{\mathcal{C}_{#1}^{'\mathrm{(eff)}}}\xspace}       % 7
\def\Ope#1    {\ensuremath{\mathcal{O}_{#1}}\xspace}                       % 2
\def\Opep#1   {\ensuremath{\mathcal{O}_{#1}^{'}}\xspace}                    % 7

\def\dkpipi     {\decay{\Dp}{\Km\pip\pip}}
\def\dpipik     {\decay{\Dp}{\pim\pip\Kp}}
\def\dkkk     {\decay{\Dp}{\Km\Kp\Kp}}
\def\dspikk     {\decay{\Dsp}{\pim\Kp\Kp}}
\def\dskkpi     {\decay{\Dsp}{\Km\Kp\pip}}
\def\dkkpi     {\decay{\Dp}{\Km\Kp\pip}}

\def\dhhh           {\decay{\Dps}{h^-_{\mathrm 1}h^+_{\mathrm 2}h^+_{\mathrm 3}}}

\newcommand{\tev}{\ifthenelse{\boolean{inbibliography}}{\ensuremath{~T\kern -0.05em eV}}{\ensuremath{\mathrm{\,Te\kern -0.1em V}}}\xspace}
\newcommand{\gev}{\ensuremath{\mathrm{\,Ge\kern -0.1em V}}\xspace}
\newcommand{\mev}{\ensuremath{\mathrm{\,Me\kern -0.1em V}}\xspace}
\newcommand{\kev}{\ensuremath{\mathrm{\,ke\kern -0.1em V}}\xspace}
\newcommand{\ev}{\ensuremath{\mathrm{\,e\kern -0.1em V}}\xspace}
\newcommand{\gevc}{\ensuremath{{\mathrm{\,Ge\kern -0.1em V\!/}c}}\xspace}
\newcommand{\mevc}{\ensuremath{{\mathrm{\,Me\kern -0.1em V\!/}c}}\xspace}
\newcommand{\gevcc}{\ensuremath{{\mathrm{\,Ge\kern -0.1em V\!/}c^2}}\xspace}
\newcommand{\gevgevcccc}{\ensuremath{{\mathrm{\,Ge\kern -0.1em V^2\!/}c^4}}\xspace}
\newcommand{\mevcc}{\ensuremath{{\mathrm{\,Me\kern -0.1em V\!/}c^2}}\xspace}

\def\mum  {\ensuremath{{\,\upmu\mathrm{m}}}\xspace}

\def\invfb   {\ensuremath{\mbox{\,fb}^{-1}}\xspace}

\newcommand{\chisq}{\ensuremath{\chi^2}\xspace}

\newcommand{\chisqip}{\ensuremath{\chi^2_{\text{IP}}}\xspace}

\def\gsim{{~\raise.15em\hbox{$>$}\kern-.85em
          \lower.35em\hbox{$\sim$}~}\xspace}
\def\lsim{{~\raise.15em\hbox{$<$}\kern-.85em
          \lower.35em\hbox{$\sim$}~}\xspace}

\def\sPlot{\mbox{\em sPlot}\xspace}
\def\sWeights{\mbox{\em sWeights}}

\def\ptot       {\mbox{$p$}\xspace}
\def\pt         {\mbox{$p_{\mathrm{ T}}$}\xspace}

\def\evtgen     {\mbox{\textsc{EvtGen}}\xspace}

\def\geant      {\mbox{\textsc{Geant4}}\xspace}

\def\photos     {\mbox{\textsc{Photos}}\xspace}

\def\pythia     {\mbox{\textsc{Pythia}}\xspace}

\def\tell1  {TELL1\xspace}
\def\ukl1   {UKL1\xspace}

\usepackage{cite} % Allows for ranges in citations
\usepackage{mciteplus}

\usepackage{longtable} % only for template; not usually to be used in PAPERs

\begin{document}

\renewcommand{\thefootnote}{\fnsymbol{footnote}}
\setcounter{footnote}{1}

%\onecolumn
% $Id: title-LHCb-PAPER.tex 111203 2017-08-08 15:28:40Z pkoppenb $
% ===============================================================================
% Purpose: LHCb-PAPER journal paper title page template
% Author: 
% Created on: 2010-09-25
% ===============================================================================

%%%%%%%%%%%%%%%%%%%%%%%%%
%%%%%  TITLE PAGE  %%%%%%
%%%%%%%%%%%%%%%%%%%%%%%%%
\begin{titlepage}
\pagenumbering{roman}

% Header ---------------------------------------------------
\vspace*{-1.5cm}
\centerline{\large EUROPEAN ORGANIZATION FOR NUCLEAR RESEARCH (CERN)}
\vspace*{1.5cm}
\noindent
\begin{tabular*}{\linewidth}{lc@{\extracolsep{\fill}}r@{\extracolsep{0pt}}}
\ifthenelse{\boolean{pdflatex}}% Logo format choice
{\vspace*{-1.5cm}\mbox{\!\!\!\includegraphics[width=.14\textwidth]{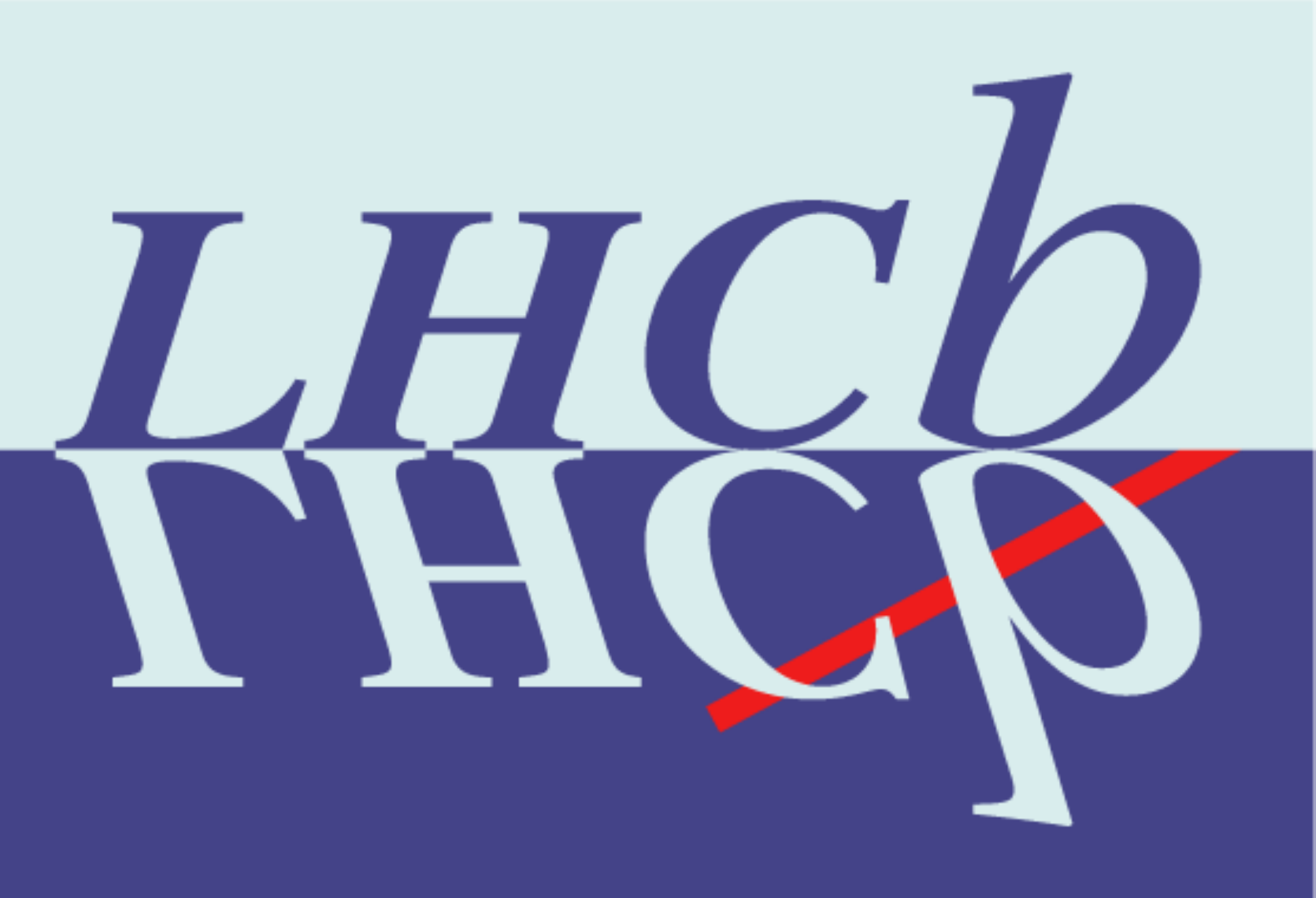}} & &}%
{\vspace*{-1.2cm}\mbox{\!\!\!\includegraphics[width=.12\textwidth]{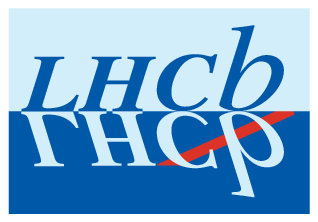}} & &}%
\\
 & & CERN-EP-2018-244 \\  % ID 
 & & LHCb-PAPER-2018-033 \\  % ID 
 & & 5 April 2019 \\ % Date - Can also hardwire e.g.: 23 March 2010
 & & \\
% not in paper \hline
\end{tabular*}

\vspace*{1.0cm}

% Title --------------------------------------------------
{\normalfont\bfseries\boldmath\huge
\begin{center}
% DO NOT EDIT HERE. Instead edit macro in main.tex to keep metadata correct
  \papertitle 
\end{center}
}

\vspace*{1.0cm}

% Authors -------------------------------------------------
\begin{center}
%In the footnote, replace 'paper' by 'Letter' in case of submission to PRL or PLB 
% Edit macro in main.tex to keep metadata correct
\paperauthors\footnote{Authors are listed at the end of this paper.}
\end{center}

\vspace{\fill}

% Abstract -----------------------------------------------
\begin{abstract}
  \noindent
The branching fractions of the doubly Cabibbo-suppressed decays  \mbox{\dkkk,}  \dpipik and \dspikk
are measured using the decays {\mbox{\dkpipi}} and \dskkpi as normalisation channels. The measurements are performed using proton-proton collision data collected with the LHCb detector at a centre-of-mass energy of 8\tev, corresponding to an integrated luminosity of 2.0\invfb. The results are
\begin{align}
 \frac {\BF(\dkkk)} {\BF(\dkpipi)}& = (6.541 \pm 0.025  \pm 0.042) \times 10^{-4},\nonumber \\
 \frac {\BF(\dpipik)} {\BF(\dkpipi)}& = (5.231 \pm 0.009 \pm 0.023)  \times 10^{-3}, \nonumber\\
\frac {\BF(\dspikk)} {\BF(\dskkpi)}& = (2.372 \pm 0.024 \pm 0.025)  \times 10^{-3},\nonumber 
\end{align}

\noindent where the uncertainties are statistical and systematic,
 respectively. These are the most precise measurements up to date.
\end{abstract}

\vspace*{1.0cm}

\begin{center}
Published in JHEP 03 (2019) 176  
\end{center}

\vspace{\fill}

{\footnotesize 
% Edit macro in main.tex to keep metadata correct
\centerline{\copyright~\papercopyright. \href{\paperlicenceurl}{\paperlicence}.}}
\vspace*{2mm}

\end{titlepage}

%%%%%%%%%%%%%%%%%%%%%%%%%%%%%%%%
%%%%%  EOD OF TITLE PAGE  %%%%%%
%%%%%%%%%%%%%%%%%%%%%%%%%%%%%%%%

%  empty page follows the title page ----
\newpage
\setcounter{page}{2}
\mbox{~}
%\newpage
%
%% Author List ----------------------------
%%  You need to get a new author list!
%\input{LHCb_authorlist.tex}
%
%The author list for journal publications is provided by the Membership Committee shortly after 'approval to go to paper' has been given.
%%It will be made available on the page
%%\verb!http://www.physik.uzh.ch/~strauman/forMemCo/LHCb-PAPER-XXXX-XXX/! .
%It will be sent to you by email shortly after a paper number has beens assigned.
%The author list should be included already at first circulation, 
%to allow new members of the collaboration to verify whether they have been included correctly.
%Occasionally a misspelled name is corrected or associated institutions become full members.
%In that case, a new author list will be sent to you.
%In case line numbering doesn't work well after including the authorlist, try moving the \verb!\bigskip! after the last author to a separate line.
%
%
%The authorship for Conference Reports should be ``The LHCb
%  collaboration'', with a footnote giving the name(s) of the contact
%  author(s), but without the full list of collaboration names.

\cleardoublepage

%\twocolumn
% %%%%%%%%%%%%% ---------

\renewcommand{\thefootnote}{\arabic{footnote}}
\setcounter{footnote}{0}

%\tableofcontents
%\cleardoublepage

\pagestyle{plain} % restore page numbers for the main text
\setcounter{page}{1}
\pagenumbering{arabic}

%\linenumbers
% $Id: introduction.tex 87303 2016-02-08 13:44:29Z lafferty $

\section{Introduction}
\label{sec:Introduction}

Precise measurements of the branching fractions of doubly Cabibbo-suppressed (DCS) decays of charmed mesons provide important information for the understanding of the decay dynamics of these particles. The theoretical description of charm-meson decays is challenging. The charm quark is not heavy enough for a reliable application of the factorisation approach and heavy-quark expansion tools, successfully used in \B-meson decays. It is also not light enough for the application of chiral  perturbation theory, as in the case of kaon decays.  Phenomenological models and approximate symmetries, such as those based on the diagrammatic approach~\cite{paper:rosner,paper:chao}, rely on the knowledge of  branching fractions and, in the case of multi-body final states, resonant structures, as key inputs. Whilst the branching fractions of some decay modes of charmed mesons are well measured, the uncertainties on branching fractions of doubly Cabibbo-suppressed decays are still large. 

In this paper, three ratios of branching fractions of DCS decays 
of \Dp and \Dsp mesons\footnote{Throughout this paper, charge conjugated decays are implied.} are measured with unprecedented precision,
 
\begin{equation}
\frac{{\cal B}(\decay{\Dp}{K^- K^+ K^+})}{{\cal B}(\decay{\Dp}{K^-\pi^+\pi^+})}, \hskip .2cm\frac{{\cal B}(\decay{\Dp}{\pi^-\pi^+K^+})}{{\cal B}(\decay{\Dp}{K^-\pi^+\pi^+})},\hskip .2cm \frac{{\cal B}(\decay{\Dsp}{\pi^-K^+K^+})}{{\cal B}(\decay{\Dsp}{K^-K^+\pi^+})}.
\end{equation}

\noindent In addition, the branching fraction of the Cabibbo-suppressed (CS) \dkkpi decay is measured relative to that of the Cabibbo-favoured (CF) \mbox{\dkpipi} decay.
\begin{figure}[hp]
\centerline{
\includegraphics[width = .35\textwidth]{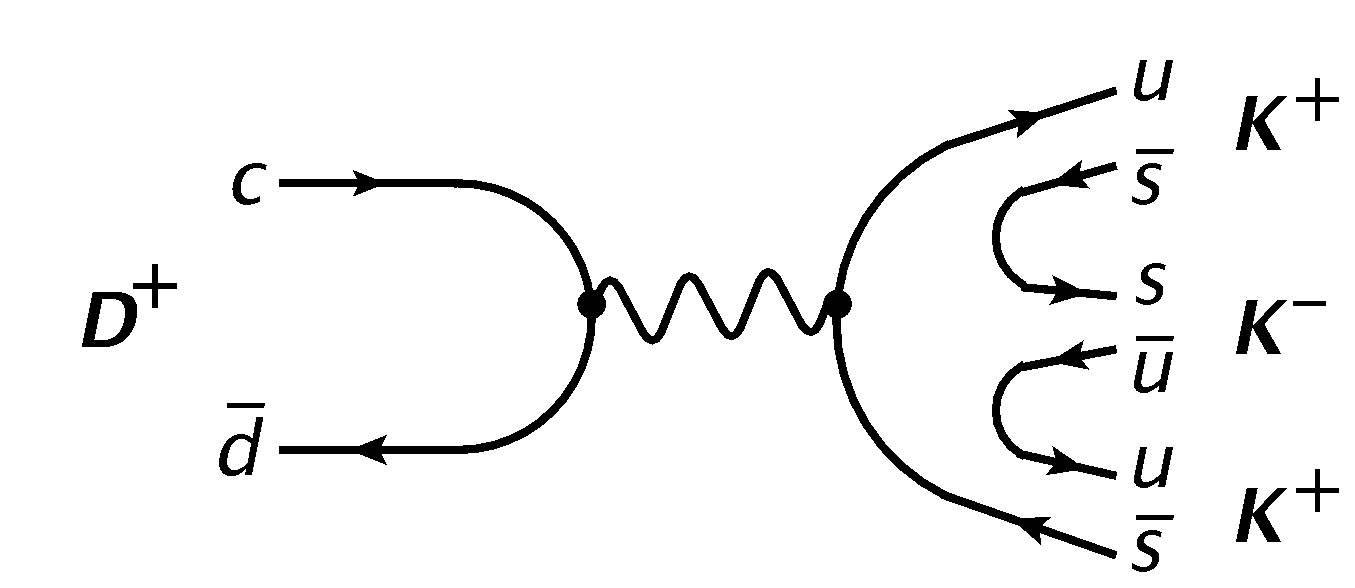}\hskip 0.3 cm
\includegraphics[width = .3\textwidth]{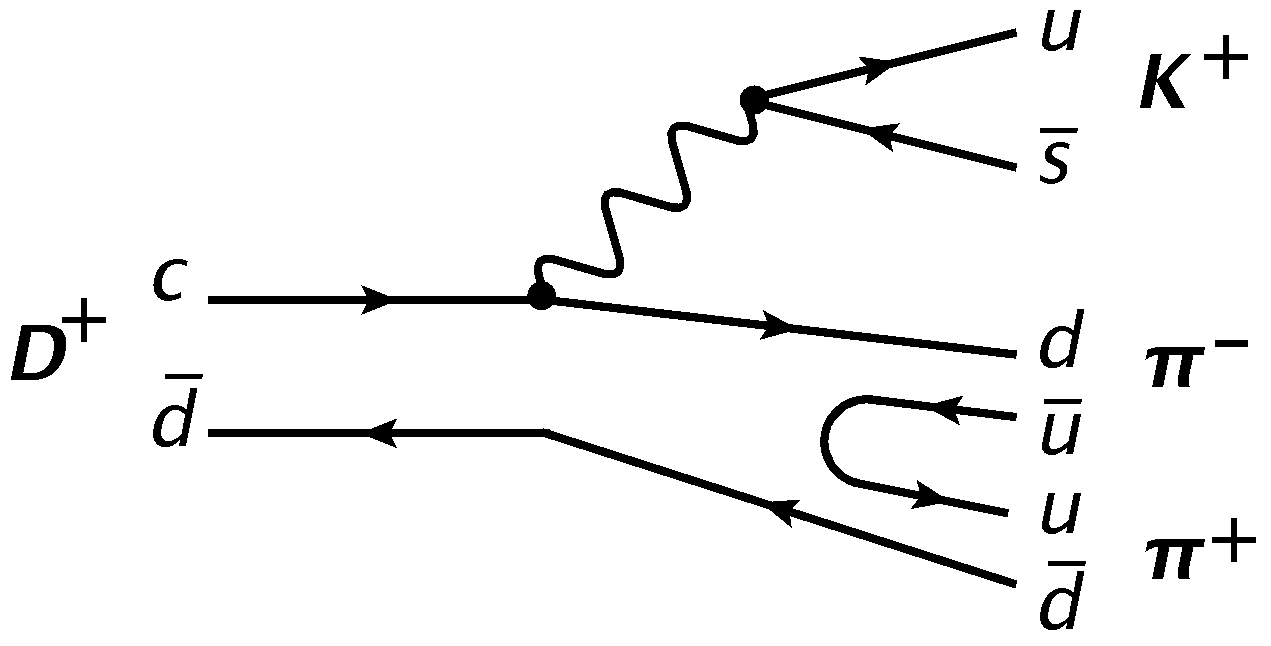}\hskip 0.3 cm
\includegraphics[width = .3\textwidth]{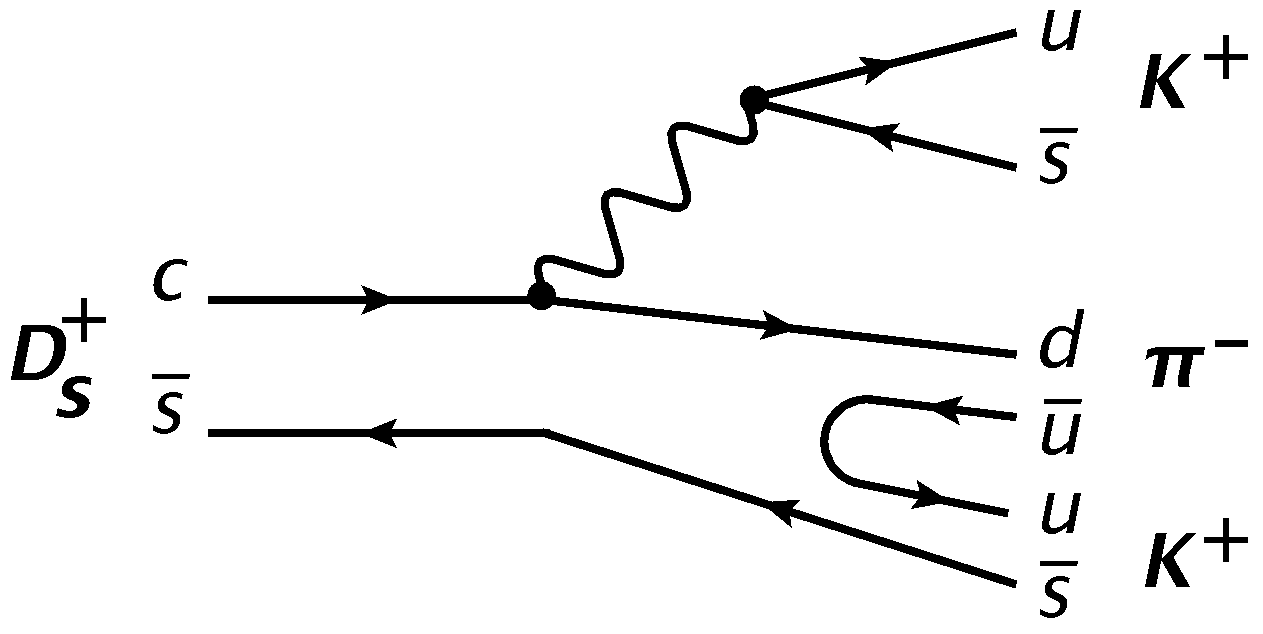}}
\centerline{ (a) \hskip 4.5 cm (b) \hskip 4.5 cm (c) }
\caption{\small Feynman diagrams for the DCS decays (a) \dkkk, (b) \dpipik and (c) \dspikk .}
\label{fig:diagrams}
\end{figure}

Tree-level diagrams for the three DCS decays are exemplified in Fig.~\ref{fig:diagrams}, where the final state particles can be produced through resonances not explicitly shown. The decay \dkkk is expected to occur through an annihilation process as in Fig.~\ref{fig:diagrams}(a) but it is also possible to produce the \Km\Kp\Kp final state through a diagram similar to that in Fig.~\ref{fig:diagrams}(b), where a \Kp\Km pair could be formed through the  \decay{\Kz\Kzb}{\Kp\Km} rescattering or through a resonance that couples to both $d\bar d$ and $s\bar s$.

The world averages~\cite{PDG2018} of these ratios of branching fractions are listed in Table~\ref{tab:wavg}. In the case of the \dkkk decay, there is only one previous measurement by the FOCUS collaboration~\cite{focuskkk},
based on a sample of 65 $\pm$ 15 decays and with a precision of 23\,\%.

\begin{table}[t]
\caption{\small World averages for the branching-fractions ratios under consideration~\cite{PDG2018}.}
\label{tab:wavg}
\centering\begin{tabular}{lr@{\,$\pm$\,}l}\hline
Ratio &  \multicolumn{2}{c}{Value [$\times 10^{-3}$]}\\\hline 
${\cal B}$(\dkkk)$/{\cal B}$(\dkpipi)   &0.95&0.22 \\
${\cal B}$(\dpipik)$/{\cal B}$(\dkpipi) & 5.77&0.22\\
${\cal B}$(\dspikk)$/{\cal B}$(\dskkpi) &2.33&0.23\\
${\cal B}$(\dkkpi)$/{\cal B}$(\dkpipi)  &  105.9&1.8      \\\hline  
\end{tabular}
\end{table}

The results presented in this paper are obtained with 
a sample of $pp$-collision data corresponding to an integrated luminosity of 2.0\,fb$^{-1}$, collected at a centre-of-mass energy of 8\tev with the LHCb detector. In Section~\ref{sec:Detector} a description of the detector and simulation is presented.
The method used to measure the ratio of branching fractions is described in Section~\ref{sec:method}. The selection is discussed in Section~\ref{sec:selection}. The determination of the efficiencies in bins of the phase space is explained in Section~\ref{sec:efficiencies}. 
The fit model and the evaluation of the signal yields are presented   in Section~\ref{sec:yields}. 
Systematic uncertainties associated with the measurements are discussed in Section~\ref{sec:systematics}. Finally, the results and conclusions are presented in Section~\ref{sec:results}.

\section{Detector and simulation}
\label{sec:Detector}

The \lhcb detector~\cite{Alves:2008zz,LHCb-DP-2014-002} is a single-arm forward
spectrometer covering the \mbox{pseudorapidity} range $2<\eta <5$,
designed for the study of particles containing \bquark or \cquark
quarks. The detector includes a high-precision tracking system
consisting of a silicon-strip vertex detector surrounding the $pp$
interaction region, a large-area silicon-strip detector located
upstream of a dipole magnet with a bending power of about
$4{\mathrm{\,Tm}}$, and three stations of silicon-strip detectors and straw
drift tubes placed downstream of the magnet.
The tracking system provides a measurement of the momentum, \ptot, of charged particles with
a relative uncertainty that varies from 0.5\% at low momentum to 1.0\% at 200\gevc.
The minimum distance of a track to a primary vertex (PV), the impact parameter (IP), 
is measured with a resolution of $(15+29/\pt)\mum$,
where \pt is the component of the momentum transverse to the beam, in\,\gevc.
Different types of charged hadrons are distinguished using information
from two ring-imaging Cherenkov detectors~\cite{LHCb-DP-2012-003}. 
Photons, electrons and hadrons are identified by a calorimeter system consisting of
scintillating-pad and preshower detectors, an electromagnetic
calorimeter and a hadronic calorimeter. Muons are identified by a
system composed of alternating layers of iron and multiwire
proportional chambers.

The polarity of the dipole magnet is reversed periodically throughout data taking. The configurations with the magnetic field upwards, \MagUp, and  downwards, \MagDown, bend respectively positively and negatively charged particles in the horizontal plane towards the centre of the LHC.

The online event selection is performed by a trigger system\cite{LHCb-DP-2012-004}, 
which consists of a hardware stage, based on information from the calorimeter and muon
systems, followed by a software stage, which applies a full event
reconstruction. 
At the hardware-trigger stage, events are required to have a muon with high \pt or a
hadron, photon or electron with high transverse energy in the calorimeters.
In the offline selection, the hardware trigger signals are associated with reconstructed particles.
Selection requirements can therefore be made on whether the decision is due to the signal candidate, other particles produced in the $pp$ collision, or a combination of both. 
The latter is used in this analysis.
The software trigger  is divided into two parts. The first part employs a partial reconstruction of the tracks, and a requirement on \pt and IP is applied to, at least, one final-state track forming the \Dps candidate. In the second part  a full event reconstruction is performed and dedicated algorithms   are used to select $D^+_{(s)}$ candidates decaying into three charged hadrons.

 In the simulation, $pp$ collisions are generated using
\pythia~\cite{Sjostrand:2006za,*Sjostrand:2007gs}  with a specific \lhcb
configuration~\cite{LHCb-PROC-2010-056}.  Decays of hadronic particles
are described by \evtgen~\cite{Lange:2001uf}, in which final-state
radiation is generated using \photos~\cite{Golonka:2005pn}. The
interaction of the generated particles with the detector, and its response, are implemented using the \geant
toolkit~\cite{Allison:2006ve, *Agostinelli:2002hh} as described in
Ref.~\cite{LHCb-PROC-2011-006}.

\section{Method}
\label{sec:method}

The ratios of branching fractions are measured as
\begin{equation}
\frac{{\cal B}(\decay{\Dps}{f_{\mathrm{signal}}})}{{\cal B}(\decay{\Dps}{f_{\mathrm{norm}}})}=\frac{N_{\mathrm{signal}}^{\mathrm{prod}}}{N_{\mathrm{norm}}^{\mathrm{prod}}},
\label{eq:brdef}
\end{equation}

\noindent where $f_{\mathrm{signal}}$ and $f_{\mathrm{norm}}$ correspond to the  final states of the signal and normalisation \Dps decays, and $N_{\mathrm{signal}}^{\mathrm{prod}}$ and  $N_{\mathrm{norm}}^{\mathrm{prod}}$ are the total number of produced  signal and normalisation decays. These numbers are determined by  correcting the observed  yields of signal ($N_{\mathrm{signal}}^{\mathrm{obs}}$) and normalisation  ($N_{\mathrm{norm}}^{\mathrm{obs}}$) decays after  full selection criteria by the total respective efficiencies, which  are obtained from simulation and from calibration data samples. Since there are no reliable decay amplitude models available for all the \dhhh decays,\footnote{Here $h$ denotes a pion or a kaon and the particle ordering is such that $h_{\mathrm{1}}$ has opposite charge with respect to the \Dps candidate.} the simulated samples  are generated according to phase space distribution. As the efficiency, $\varepsilon$, is not uniform across the phase space, both the efficiency and the number of observed decays are obtained in bins of the Dalitz plot (DP)~\cite{Dalitz:1953cp}, built with two independent invariant masses squared, denoted as $s(h_{\mathrm{1}}^- h_{\mathrm{2}}^+)$ and $s(h_{\mathrm{1}}^-h_{\mathrm {3}}^+)$. The total number of produced decays is then evaluated as

\begin{equation}
N^{\mathrm{prod}}=\sum_{i}^{N_{\mathrm{bins}}}\frac{N_{i}^{\mathrm{obs}}}{\varepsilon_{i}}, 
\label{eq:nprod}
\end{equation}
where the index $i$ runs over the bins within the kinematically allowed region of the decay DP.   
  When the decay has two identical particles in the final state, the DP is folded, with axes corresponding to the highest and lowest values of the two invariants, $s_{\mathrm{hi}}(h^-h^{\prime+})$ and $s_{\mathrm{lo}}(h^-h^{\prime+})$.

The distributions of both the efficiencies and observed yields over the phase space are obtained separately for statistically independent datasets split by magnet polarity. For each pair of signal and normalisation decays, the final experimental result is the combination of the \MagDown and \MagUp measurements of the ratio of branching fractions.

Systematic uncertainties are estimated using the ratios of observed yields $N_{\mathrm{signal}}^{\mathrm{obs}}/N_{\mathrm{norm}}^{\mathrm{obs}}$ and the ratios of average efficiencies, where the average is over the DP bins with weights given by the corresponding yields of observed candidates. They are also obtained separately for the different magnet polarities. The contributions from the relative uncertainties on the ratios of yields and on effective efficiencies are then added in quadrature to provide  the relative uncertainty on each ratio of branching fractions.

\section{Offline selection}
\label{sec:selection}

The offline candidate selection reduces the combinatorial background and suppresses specific peaking structures in the various mass spectra. These structures are due to crossfeeds from decays of other charm particles, which occur when one or more final-state particles are misidentified or not reconstructed.  

A first set of requirements exploits the decay topology by selecting combinations of three charged hadrons forming a good quality decay vertex, well detached from the PV. The PV is that with the smallest value of \chisqip, where \chisqip\ is defined as the difference in the vertex-fit \chisq of the PV reconstructed with and without the particle under consideration, in this case the \Dps candidate. The requirements at this level are made on the following quantities: the distance between the PV and the \Dps decay vertex; the IP of the \Dps candidate; the angle between the reconstructed \Dps momentum and the flight direction; the $\chi^2$ of the \Dps decay vertex fit; the distance of closest approach between any two final-state tracks; and the momentum, the transverse momentum and the \chisqip of the \Dps candidate and of its decay products. For each branching-fraction ratio measurement, signal and normalisation-channel candidates are  selected with the same topology requirements, allowing a partial cancellation of the systematic uncertainties. 
Besides being effective to reduce combinatorial background, these topology criteria suppress the background from the decays $D^{*+}\to D^0\pi^+$, where the $D^0$ decays to two charged hadrons, such as $D^0\to K^-\pi^+$ or $D^0\to K^-\pi^+\pi^0$.

Particle identification (PID) criteria are used to distinguish between kaons and pions and to veto muons from semileptonic decays with two charged hadrons and a muon in the final state, such as the $\D^+ \to K^-\pi^+\mu^+\nu_{\mu}$ decay. Further selection criteria based on more stringent PID requirements or invariant-mass vetoes are used to suppress crossfeeds contributing to each decay mode, except for the \dspikk channel, which does not present this kind of contamination.

The two main crossfeeds in the \dkkk channel are those from \Lc decays into 
$K^-K^+p$  and $K^-p \pi^+$ final states. The former is the dominant contribution, in spite of being
Cabibbo suppressed, since this background is caused by a single $p-K$ misidentification. 
 These backgrounds are removed using invariant-mass vetoes. Candidates
are reconstructed under the $K^-K^+p$  and $K^-p \pi^+$ mass hypotheses and rejected if the
resulting invariant masses are within [2280, 2296]\mevcc. This veto is slightly different for other decay modes as the reconstructed width of the \Lc mass peak is affected by the decay channel-dependent selection criteria.

The main exclusive backgrounds for the   \dpipik decay are the 
fully reconstructed decays \dskkpi , \dkpipi , $\Lc\to\pi^-\pi^+p$ and 
\mbox{$D^+ \to \KS K^+$}, where \KS decays to \pim\pip.
 The \dskkpi decay is the most abundant contamination, occurring
when the $K^+$ meson is misidentified as a pion. The contamination from the decay \dkpipi is due to a double $K-\pi$ misidentification. These two backgrounds are suppressed by
stringent PID requirements on the kaon and opposite-charge pion candidates. 
The crossfeed from the  $\Lc\to\pi^-\pi^+p$  decays, on the other hand, is eliminated by an invariant-mass veto. The $p-K$ misidentification occurs mostly at high momenta, where the discrimination between these two particles is limited. 
Candidates are reconstructed under the $\pi^-\pi^+ p$ hypothesis and rejected if their invariant mass is within the interval \mbox{[2275, 2300]\mevcc}. 
The decay  $D^+ \to \KS(\pi^-\pi^+)K^+$ has the same final state as the \dpipik
decay, hence this contamination cannot be suppressed using PID. In this case, candidates with
 $\pi^-\pi^+$ invariant mass within the interval [488, 508]\mevcc are discarded.

The main backgrounds in the \dskkpi sample are the \dkpipi and the $\Lc\to K^-p \pi^+$ decays.
A stringent PID requirement on the $K^+$ candidate is used to suppress the contamination
from \dkpipi  decays, whereas an invariant-mass veto eliminates the $\Lc\to K^-p \pi^+$ background.
The $K^-K^+\pi^+$ candidate is reconstructed as $K^-p\pi^+$  and  the candidate is discarded if the resulting invariant mass
is within \mbox{[2275, 2305]\mevcc}.

The $\Lc\to K^- p \pi^+$  decay is the main specific background contribution  in the \mbox{\dkkpi} sample. The  $K^-K^+\pi^+$  candidates are reconstructed as $K^- p \pi^+$ and those with invariant mass within \mbox{[2275, 2305]\mevcc} are vetoed.

There are two backgrounds in the  \dkpipi sample, the decays \mbox{\dskkpi} and 
$\Lc \to   K^- p \pi^+$. To reject the \Lc background the $K^-\pi^+\pi^+$  candidates are reconstructed as $K^- p \pi^+$ and those with invariant mass within \mbox{[2280, 2300]\mevcc} are vetoed.
The crossfeed from \dskkpi is suppressed using a stringent PID requirement on the pion candidate with the highest momentum.

\section{Efficiencies}
\label{sec:efficiencies}

In order to take into account the variation of the efficiencies across the phase space, the measurement of the ratios of branching fractions in this analysis is based upon the correction of the observed yields in bins of the corresponding DP.

In each bin $i$ of the DP, the overall selection efficiencies for signal and normalisation modes, $\varepsilon_{i}$ in Eq.~\ref{eq:nprod}, are factorised into components that are independently measured. The acceptance due to the detector geometry and the efficiencies due to trigger, final state particles reconstruction, offline selection and invariant-mass vetoes are obtained from simulation.  

The PID efficiency of each candidate is estimated by multiplying the  efficiencies for each final-state particle, which are evaluated from calibration samples of $\Dz\to\Km\pip$ decays~\cite{LHCb-PUB-2016-021} and depend on the particle momentum, pseudorapidity and event charged-particle multiplicity. Average PID efficiencies are in the range of 60 to 70\%. 

There are some small differences in the hardware trigger and tracking efficiencies between data and simulation. These differences are accounted for by weighting the simulation using data. The tracking-correction weight is obtained by multiplying the weights for each final-state particle, determined as a function of the particle momentum, transverse momentum, dipole magnet polarity and event charged-particle multiplicity~\cite{LHCb-DP-2013-002}. The impact of this correction on the individual efficiencies is at the level of 3\%. 

The trigger efficiency correction follows the method described in Ref.~\cite{LHCb-PAPER-2014-036}. The total data sample for each decay mode is separated into two mutually exclusive subsamples. The first is composed of candidates that are triggered at the hardware level  by one or more of the final state particles interacting  in the hadronic calorimeter. The second is composed of candidates triggered only by particles in the rest of the event. The correction makes use of calibration data samples of $\Dz\to\Km\pip$ decays and affects differently these two subsamples. The correction factors are evaluated as a function of the DP position for each subsample and combined into a single efficiency correction map according to their proportions in data.

\begin{figure}[hpb!]
\begin{center}
\includegraphics*[width=.44\textwidth]{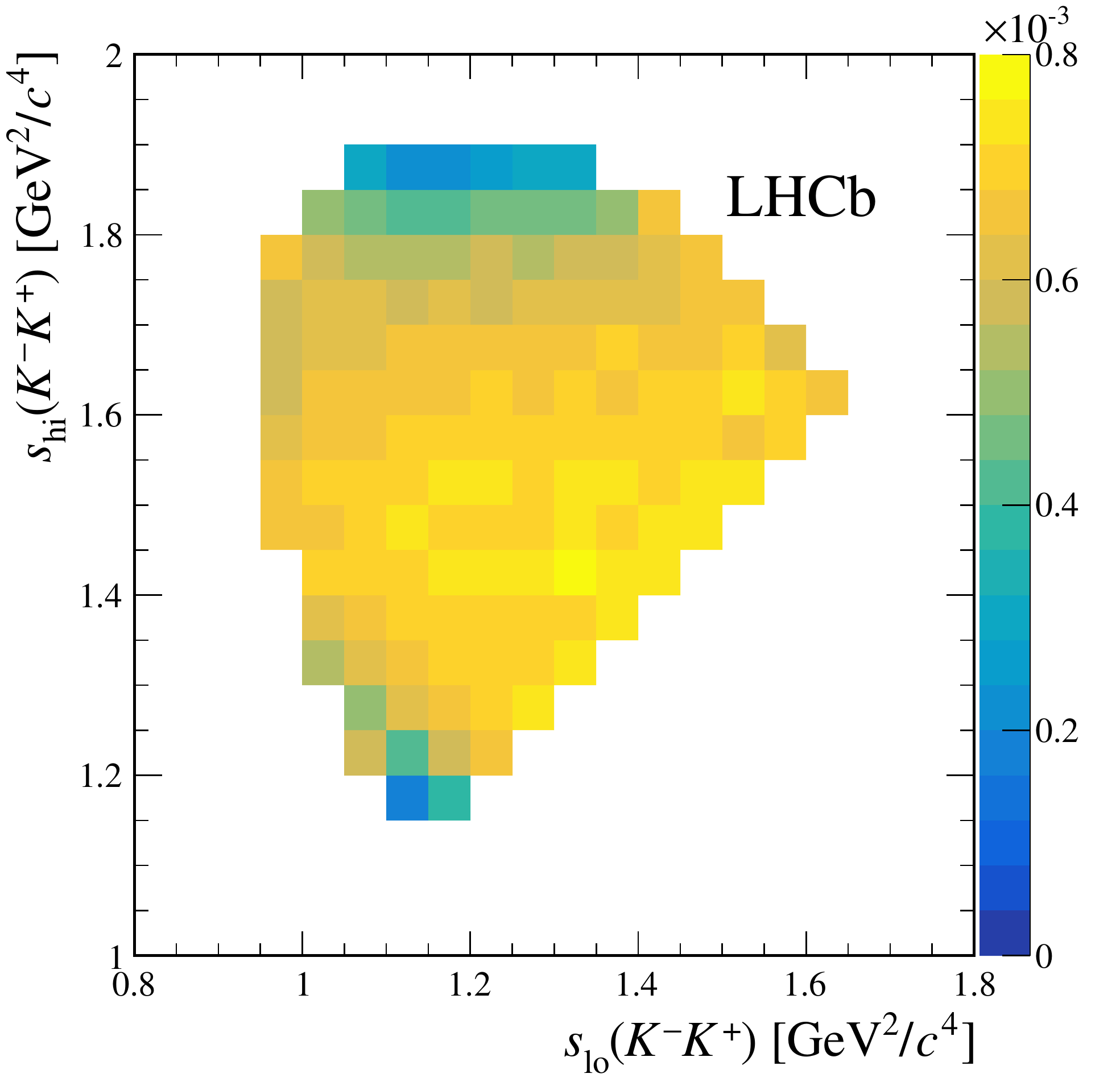}
\includegraphics*[width=.44\textwidth]{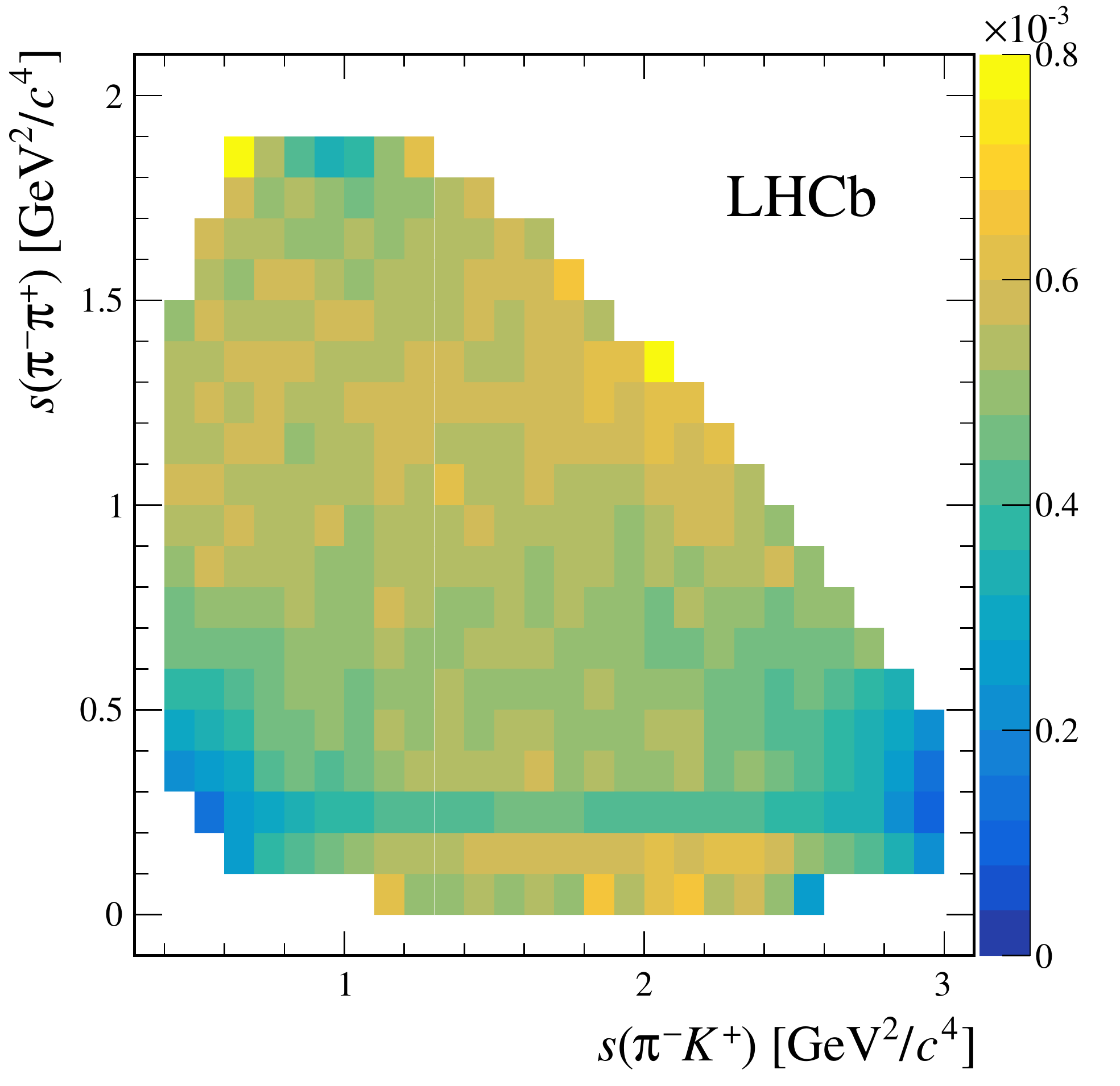}
\includegraphics*[width=.44\textwidth]{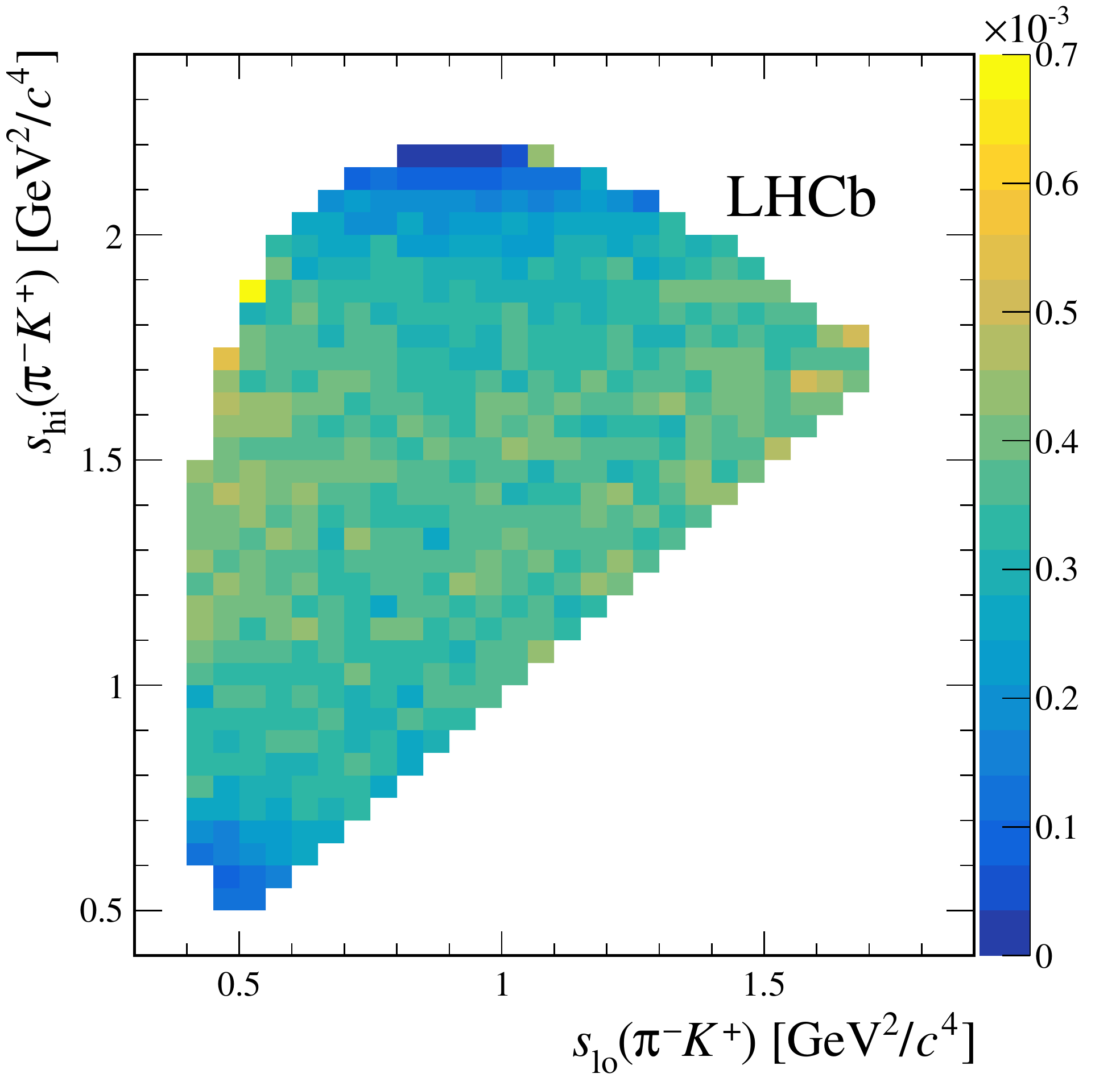}
\includegraphics*[width=.44\textwidth]{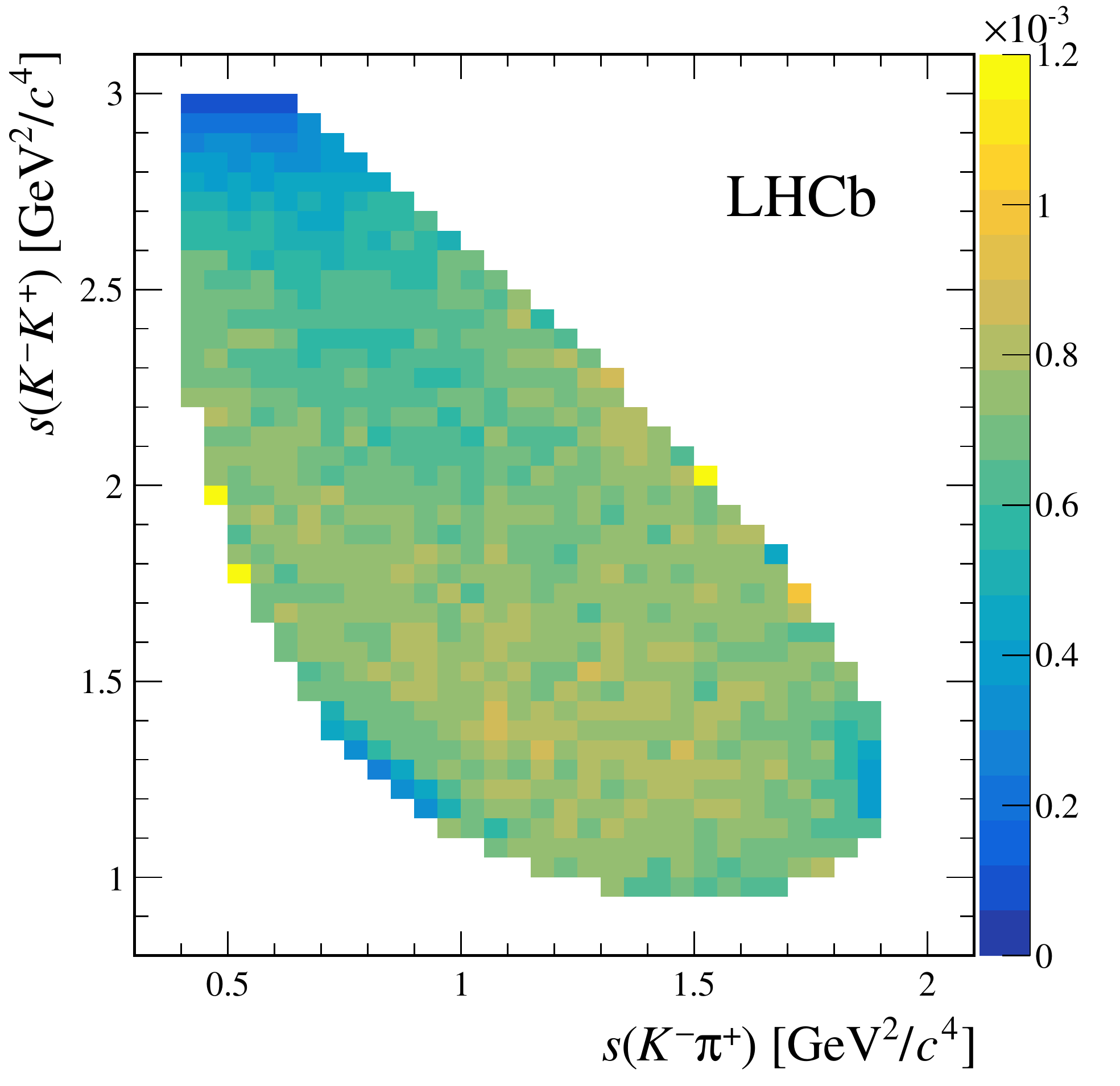}
\includegraphics*[width=.44\textwidth]{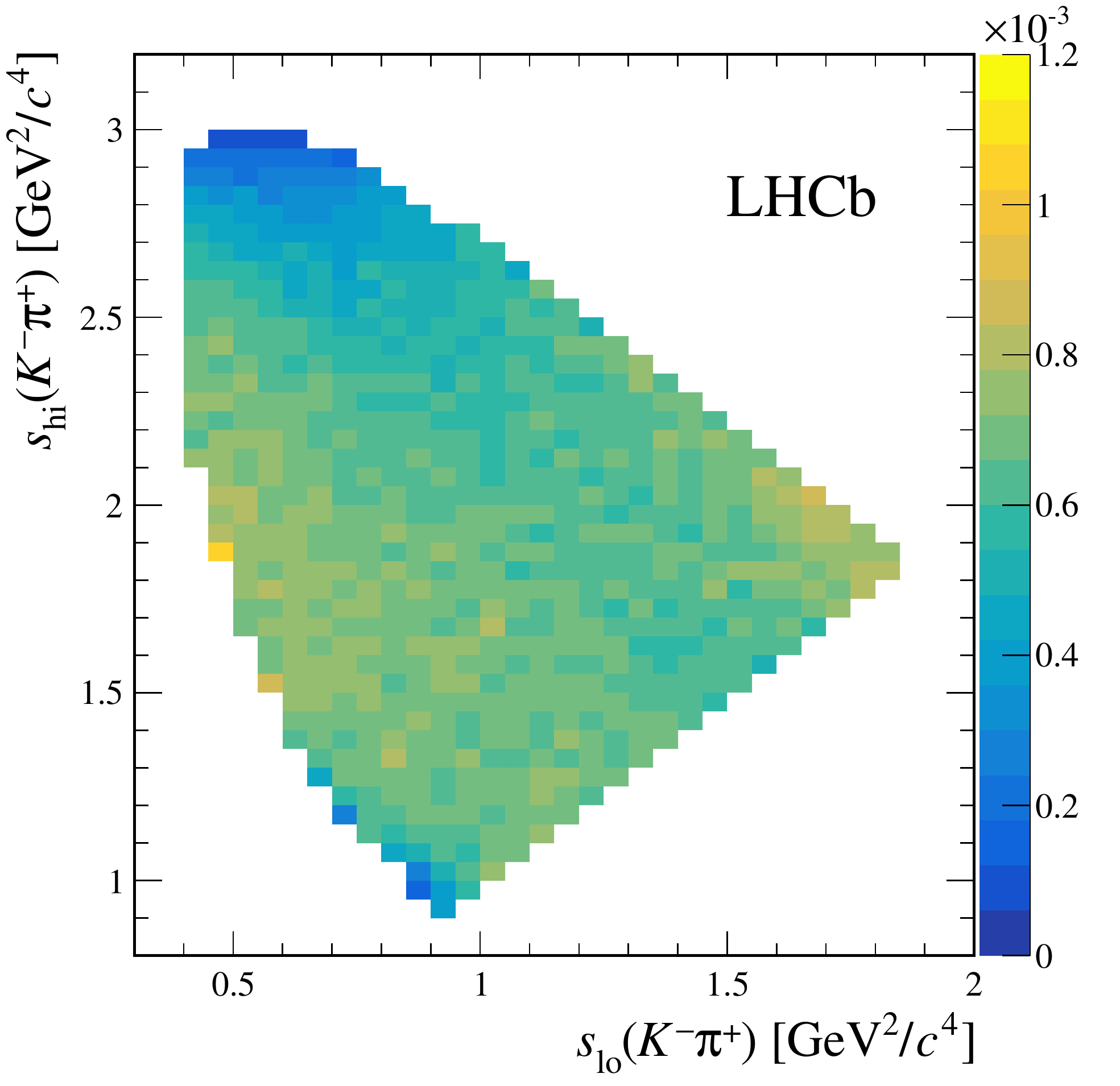}
\includegraphics*[width=.44\textwidth]{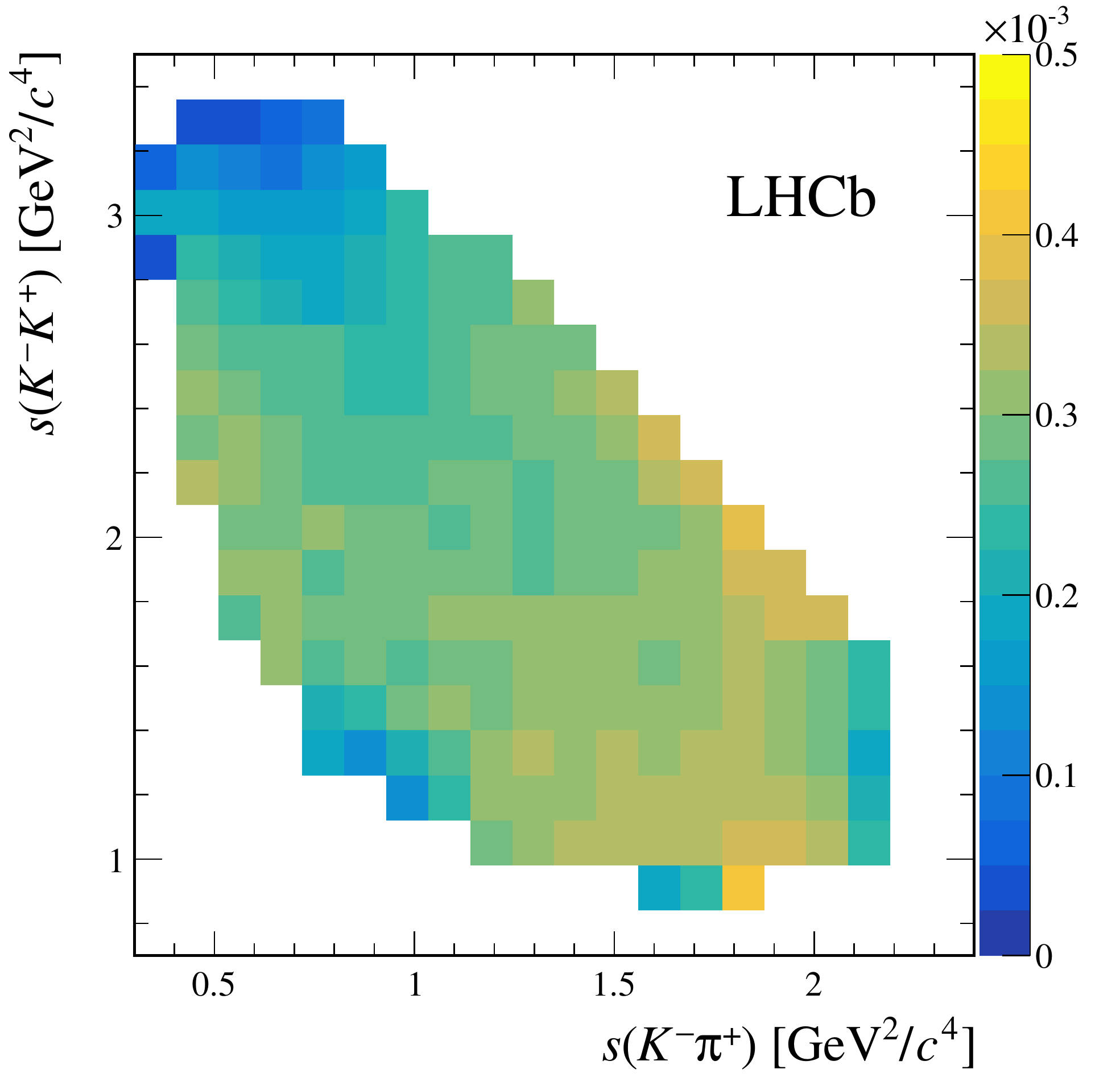}
\end{center}
    \vspace*{-.5cm}
\caption{Efficiency maps for (top left) \dkkk, (top right) \dpipik,  (middle left)  \dspikk, (middle right)  \dkkpi,  (bottom left) \dkpipi and (bottom right) \dskkpi decays with PID efficiency, tracking, multiplicity and hardware trigger efficiency corrections, for \MagDown polarity.}
  \label{fig:MCDalitz}
\end{figure}

The final efficiency maps, obtained after the full selection and corrections described above, are shown in Fig.~\ref{fig:MCDalitz}, for all decays, for \MagDown polarity (the plots for \MagUp are similar). The binning schemes used for each mode are introduced in these plots. 
The corresponding average efficiencies vary among the different decay modes from $2.7\times 10^{-4}$ (for \mbox{\dskkpi}) to $7.0\times 10^{-4}$ (for \mbox{\dkkpi}). The different lifetimes of the parent mesons and different PID criteria are the predominant contributions to this variation. The ratios between signal and normalisation channels are given in Table~\ref{tab:Reff}. The impact of the different corrections (tracking, trigger and charged-particle multiplicity) applied to the efficiencies is below 1.5\% for all ratios. 
 
\begin{table}[pt!]
 \caption{\small Ratios of average efficiencies for the full selection. The quoted uncertainty is due to the limited size of the simulated sample only (see Section~\ref{sec:systematics}).}
 \vspace*{-0.5cm}
  \begin{center}
   \begin{tabular}{lr@{\,$\pm$\,}lr@{\,$\pm$\,}l}\hline
   Ratio of efficiencies   & \multicolumn{2}{c}{\MagDown}           & \multicolumn{2}{c}{\MagUp}   \\\hline
${\varepsilon_{\dkkk}}/{\varepsilon_{\dkpipi}}$  & 1.0024 & 0.0034 &   1.0077 &  0.0033  \\   
${\varepsilon_{\dpipik}}/{\varepsilon_{\dkpipi}}$ & 0.958 &  0.005 &   0.956 &  0.005  \\  
${\varepsilon_{\dspikk}}/{\varepsilon_{\dskkpi}}$& 1.242  &  0.013  &   1.215 &  0.014  \\    
${\varepsilon_{\dkkpi}}/{\varepsilon_{\dkpipi}}$& 1.096 &  0.008 &   1.108 &  0.009  \\    \hline
   \end{tabular}
   \end{center}
   \label{tab:Reff}
   \end{table}

\vspace*{-0.5cm}
\section{Determination of the yields}
\label{sec:yields}

The yields of the signal and normalisation channels are determined
by extended binned maximum-likelihood fits to the invariant-mass distribution of each sample independently.
For each channel, the signal probability distribution function (PDF) 
is represented by a sum of a Gaussian function and two Crystal Ball (CB)~\cite{Skwarnicki:1986xj} functions, while the background is modelled by
an exponential function. The signal PDF is

\begin{multline}
{\cal P}_{\rm sig}(M) = f_{\rm G} \times {\rm G}(\mu, \sigma_{\rm G}) + (1-f_{\rm G})\times
\left[ f_{\rm CB} \times {\rm CB}_1(\mu, R_{1}\sigma_{\rm G},\alpha_1,N_1) + \right. \\
\left.  (1-  f_{\rm CB})\times {\rm CB}_2(\mu, R_{2}\sigma_{\rm G},\alpha_2,N_2)\right], 
\label{spdf}
\end{multline}

\noindent where $\mu$ and $\sigma_{\rm G}$ are the mean value and the width of the Gaussian function G. The two Crystal Ball functions, CB$_1$
and CB$_2$, have widths  $R_{1}\sigma_{\rm G}$ and  $R_{2}\sigma_{\rm G}$, and tail parameters $\alpha_1$, $N_1$ and $\alpha_2$,  $N_2$. A common 
parameter, $\mu$, describes the most probable mass value of the two Crystal Balls and the mean of the Gaussian function. 

The fractions of each PDF component are $f_{\rm G}$ for the Gaussian function, $(1-f_{\rm G})\times f_{\rm CB}$ for CB$_1$ and $(1-f_{\rm G})\times(1-f_{\rm CB})$ for CB$_2$.
The parameters $\alpha_i$, $N_i$, $R_{i}$, $f_{\rm CB}$ and $f_{\rm G}$ defining the signal PDF are fixed to the values obtained from a fit to the simulation sample. 
The position of the signal mass peak presents a small dependence on the charge of the \Dps meson and on the magnet polarity. Therefore, the samples are divided into four subsamples to ensure a precise determination of the yields.

\begin{figure}[t]
\begin{center}
\includegraphics*[width=.49\textwidth]{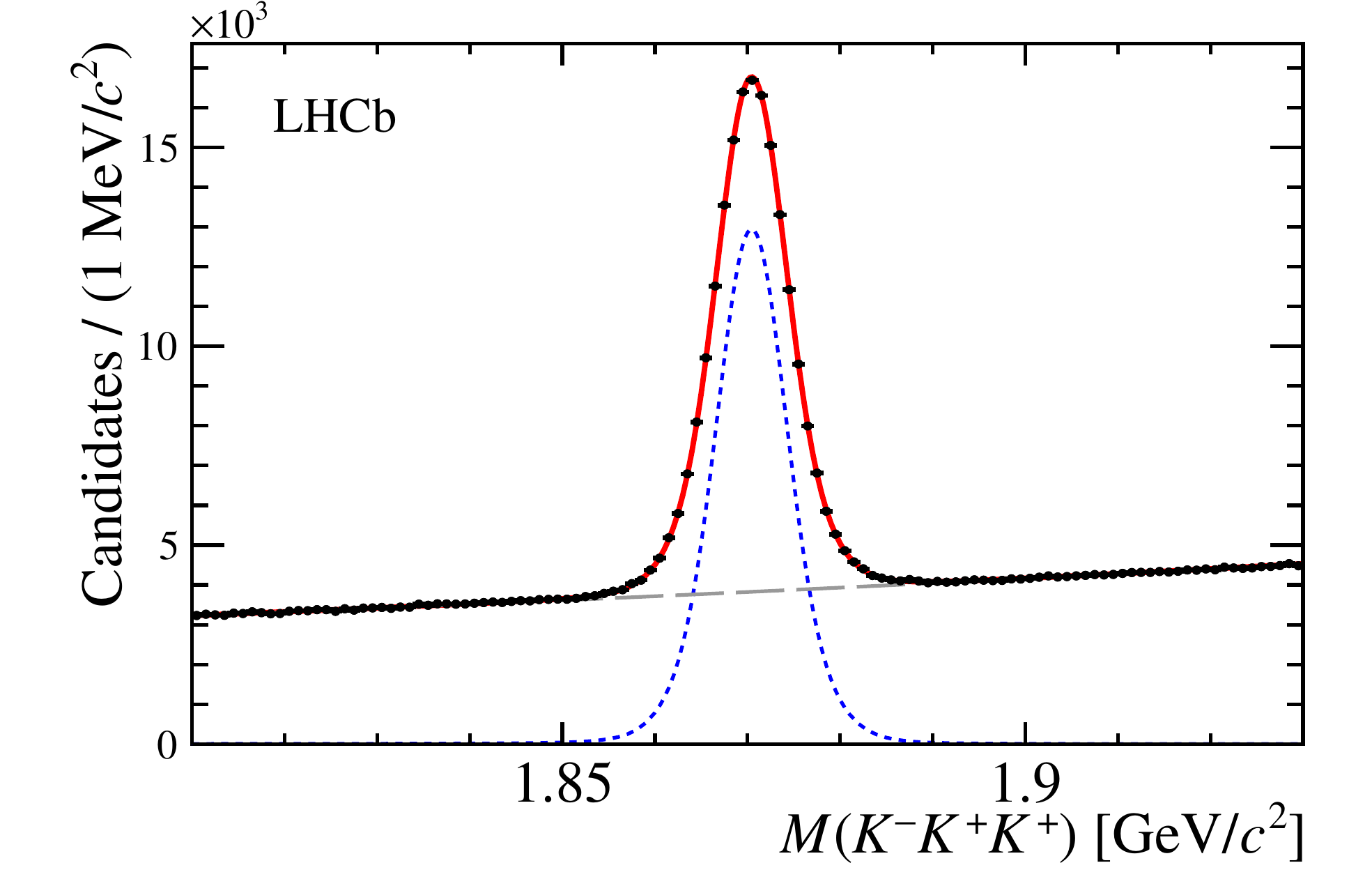}
\includegraphics*[width=.49\textwidth]{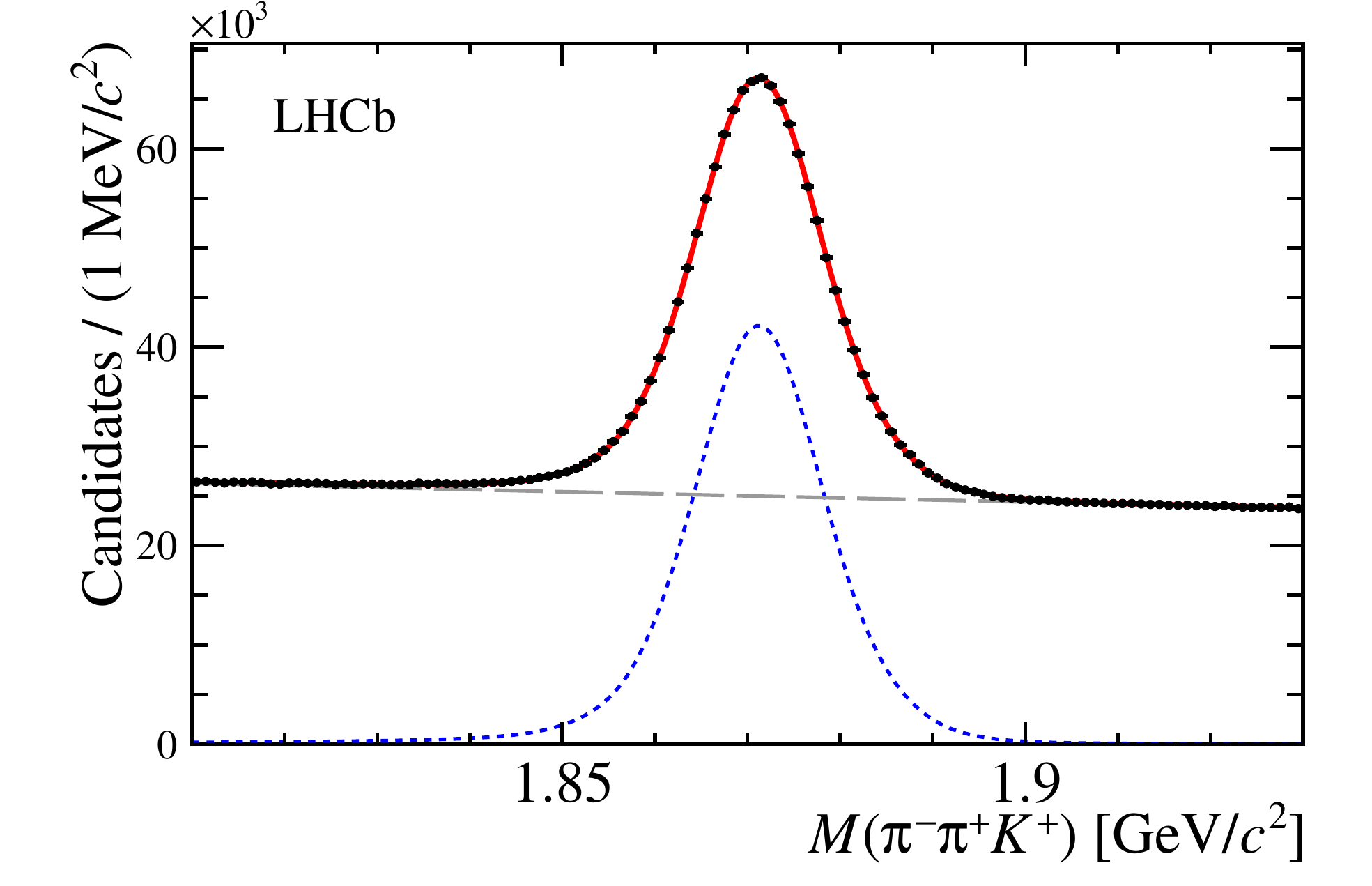}
\includegraphics*[width=.49\textwidth]{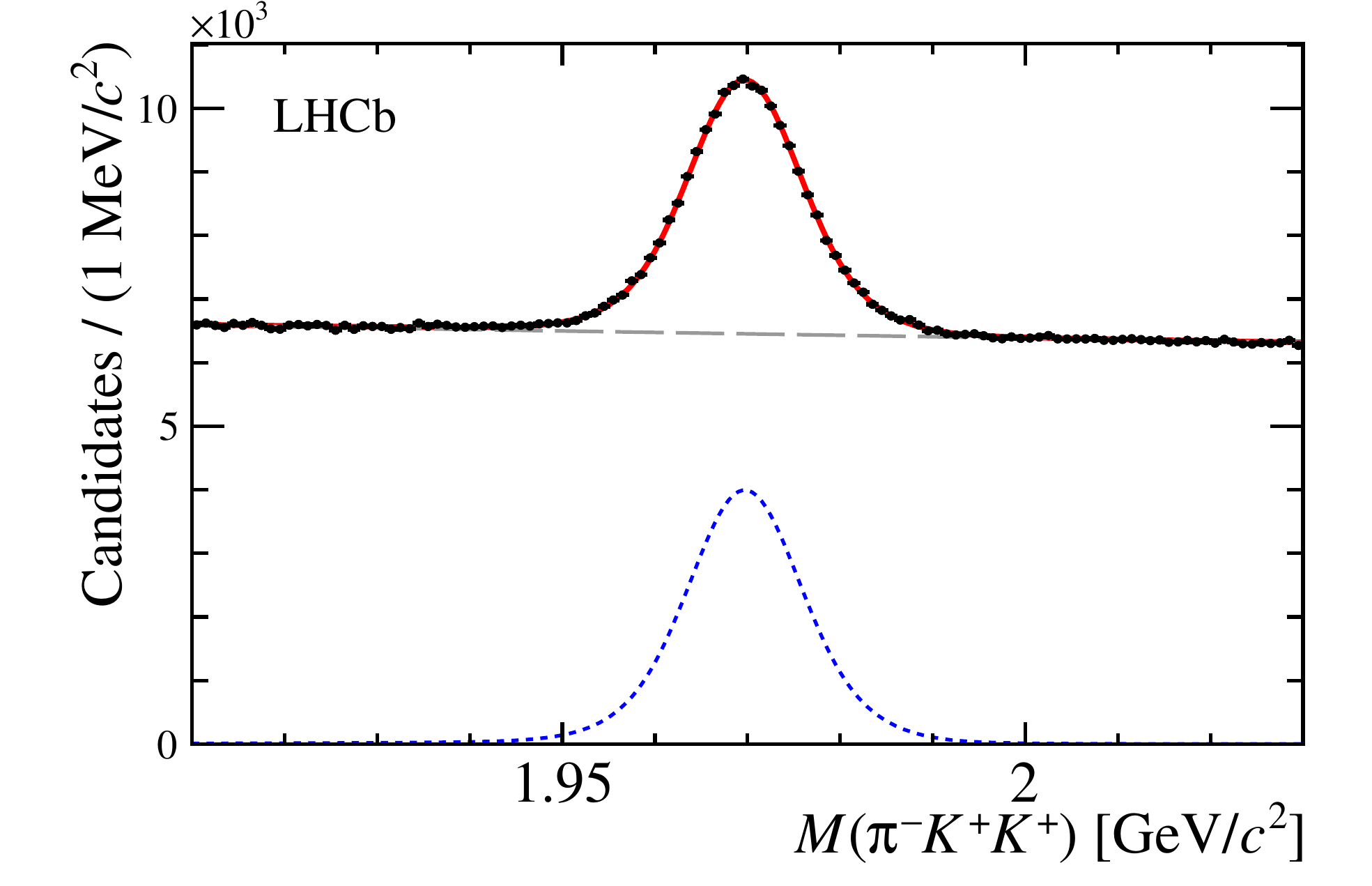}
\includegraphics*[width=.49\textwidth]{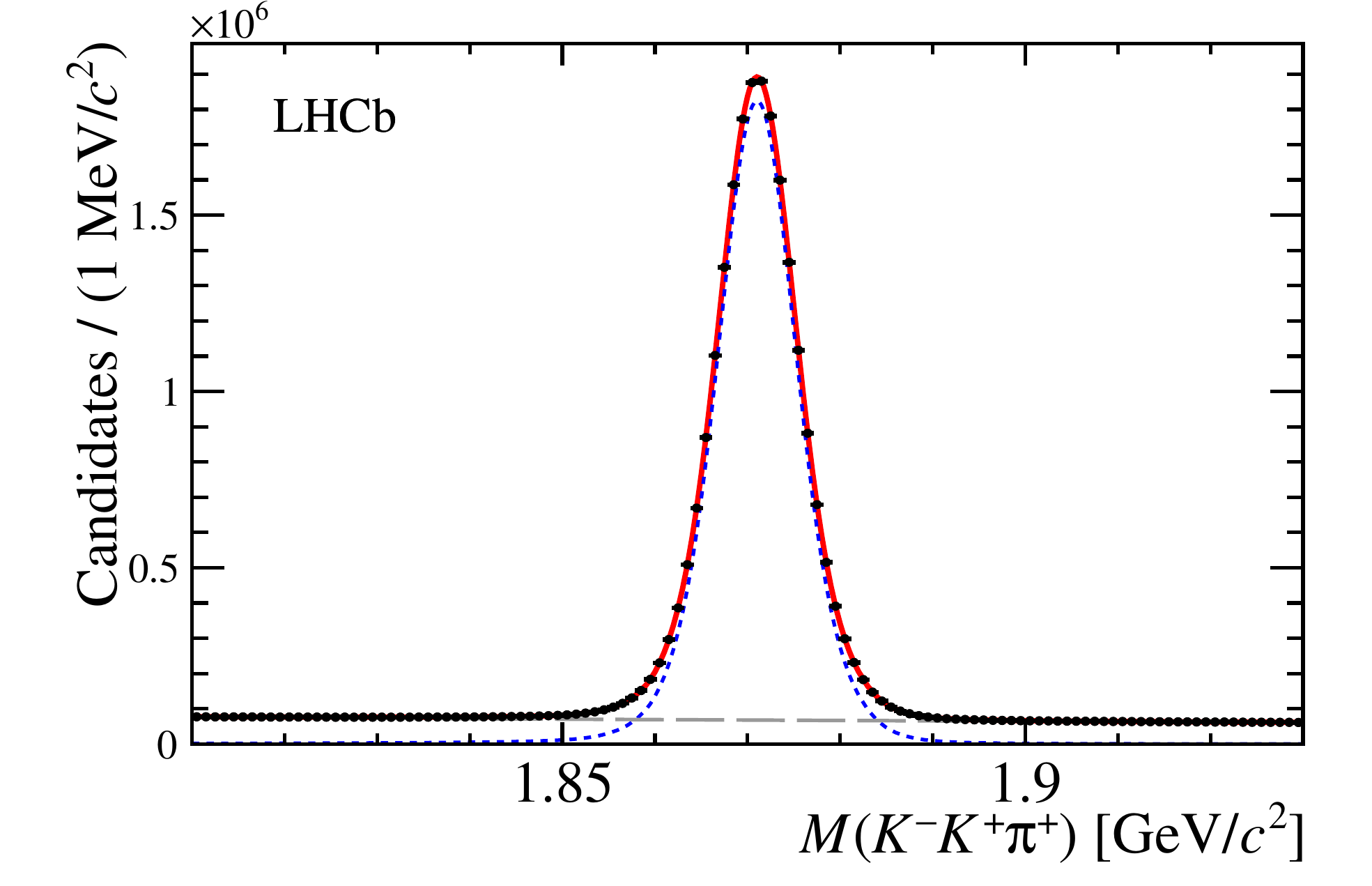}
\end{center}
    \vspace*{-.7cm}
\caption{\small  Invariant-mass distributions of (top left) \dkkk, (top right) \dpipik, (bottom left) \dspikk  and (bottom right) \dkkpi  with the corresponding fit result superimposed (red solid line). The blue dotted line corresponds to the signal PDF and the dashed grey line shows the background PDF.}
  \label{fig:signalfits}
\end{figure}

\begin{figure}[t]
\begin{center}
\includegraphics*[width=.49\textwidth]{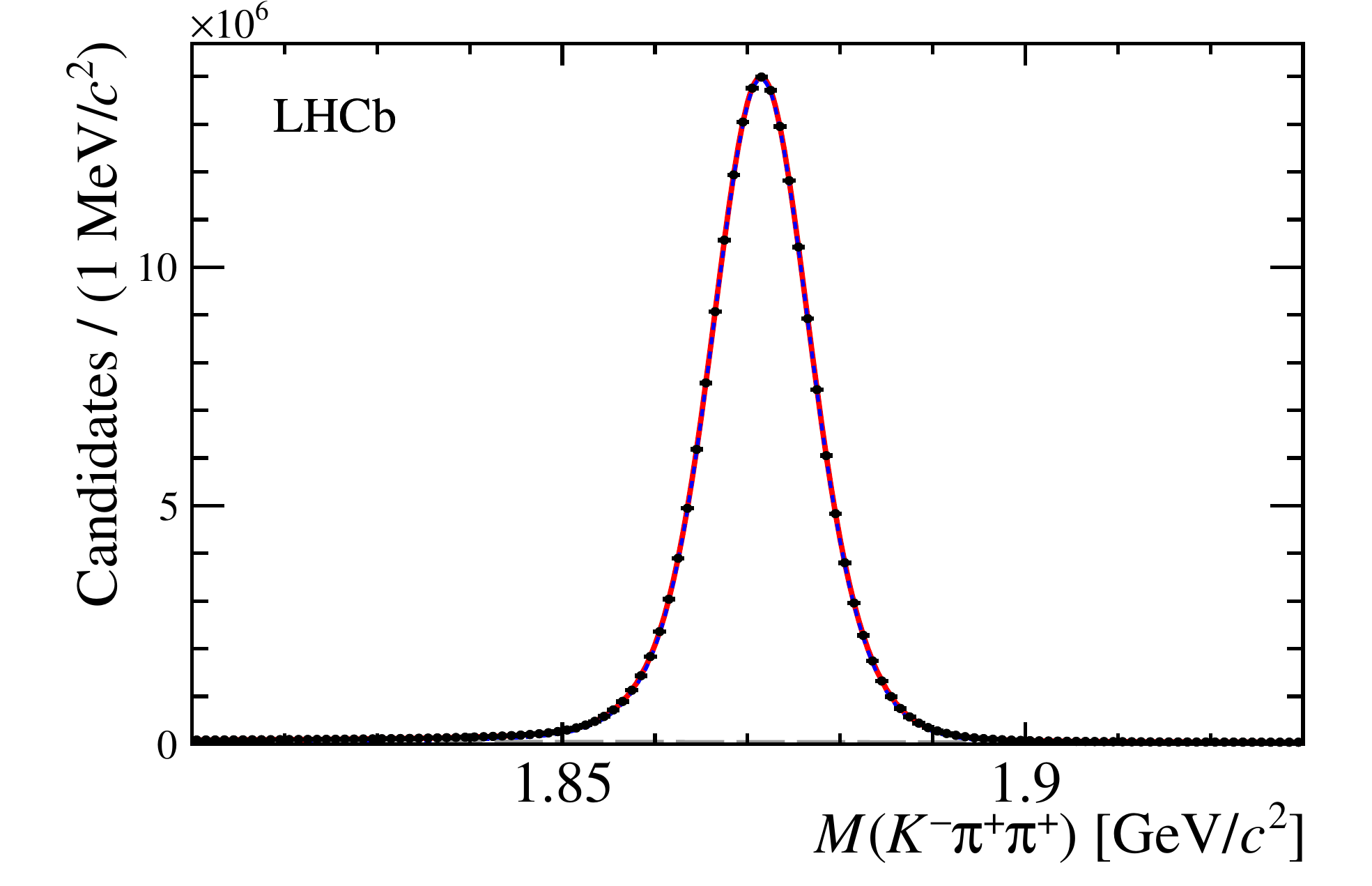}
\includegraphics*[width=.49\textwidth]{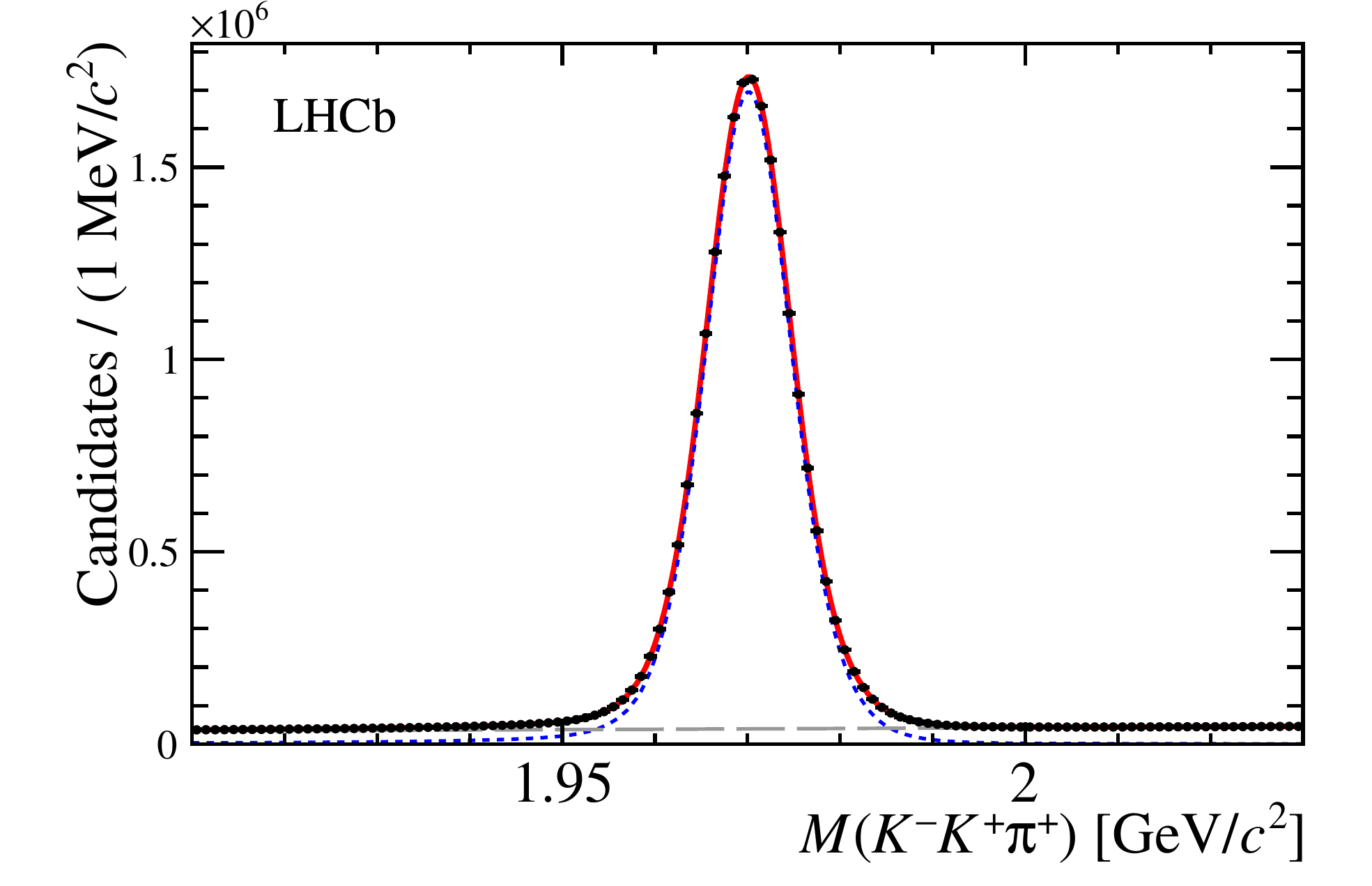}
\end{center}
    \vspace*{-.7cm}
\caption{\small Invariant-mass distributions of candidates of the normalisation modes (left) \mbox{\dkpipi}  and (right) \dskkpi  with the corresponding fit result superimposed (red solid line). The blue dotted line corresponds to the signal PDF and the dashed grey line shows the background PDF.}
  \label{fig:normfits}
\end{figure}

Due to the large size of the samples of CF channels \mbox{\dkpipi}, \dskkpi and for the CS channel \mbox{\dkkpi}, the convergence and goodness of the fit  are sensitive to the  momentum-dependent resolution of the \Dps candidate, making it difficult  to fit these samples with a single set of parameters. For this reason, the \dkpipi and \dkkpi (\dskkpi) samples  are further divided into 50 (20) bins of \Dps momentum. The variation of $\sigma_{\rm G}$ over the momentum bins is of the order of 50\%. For each magnet polarity, the total signal yield, shown in Table~\ref{tab:DataYields}, is the sum of the yields in the different subsets. For illustration purposes, the invariant-mass distribution for each of the DCS decay modes and for the CS channel are shown in Fig.~\ref{fig:signalfits} for the whole sample, summing also over the two magnet polarities, with the associated fit results superimposed. The mass distributions for the CF normalisation modes are shown in Fig.~\ref{fig:normfits}. 

\begin{table}[pt!]
\caption{\small Observed yields for signal and normalisation modes with statistical uncertainties. The entry for the decay \mbox{\dkpipi $^{(\dagger)}$} corresponds to the yields obtained from the fit to  \dkpipi sample with cuts optimised for the  \dkkk and \dkkpi selections. The entry  for the decay \dkpipi $^{(\dagger\dagger)}$ is for fits to \dkpipi samples with cuts optimised for the \dpipik selection.}
 \vspace*{-0.5cm}
\begin{center}
\begin{tabular}{l r@{\,$\pm$\,}l r@{\,$\pm$\,}l r@{\,$\pm$\,}l} \hline
 & \multicolumn{6}{c}{Yields [$\times 10^3$]}    \\
     Channel      & \multicolumn{2}{c}{\MagDown} & \multicolumn{2}{c}{\MagUp} & \multicolumn{2}{c}{Total}\\\hline
\dkkk                 & 67.61       & 0.33     & 66.69    & 0.33  & 134.30   & 0.47  \\
\dpipik               & 401.2       & 1.0      & 393.7   & 1.0  & 794.9    & 1.4  \\
\dspikk               & 33.7       & 0.4     & 33.6    & 0.4  & 67.2    & 0.5  \\
\dkkpi                & 11\,657     & 4      & 11\,482  & 4   & 23\,139  & 5  \\
\dkpipi $^{(\dagger)}$& 103\,282 &  10       & 101\,008   & 10    & 204\,290   & 14  \\
\dkpipi $^{(\dagger\dagger)}$ & 80\,197&  10      & 78\,530  & 10   & 158\,727   & 13   \\
\dskkpi            &11\,629   & 4      & 11\,414  & 4   & 23\,044  & 5  \\\hline
\end{tabular}
\end{center}
\label{tab:DataYields}
\end{table}

\begin{figure}[hbtp]
\begin{center}
\includegraphics*[width=.47\textwidth]{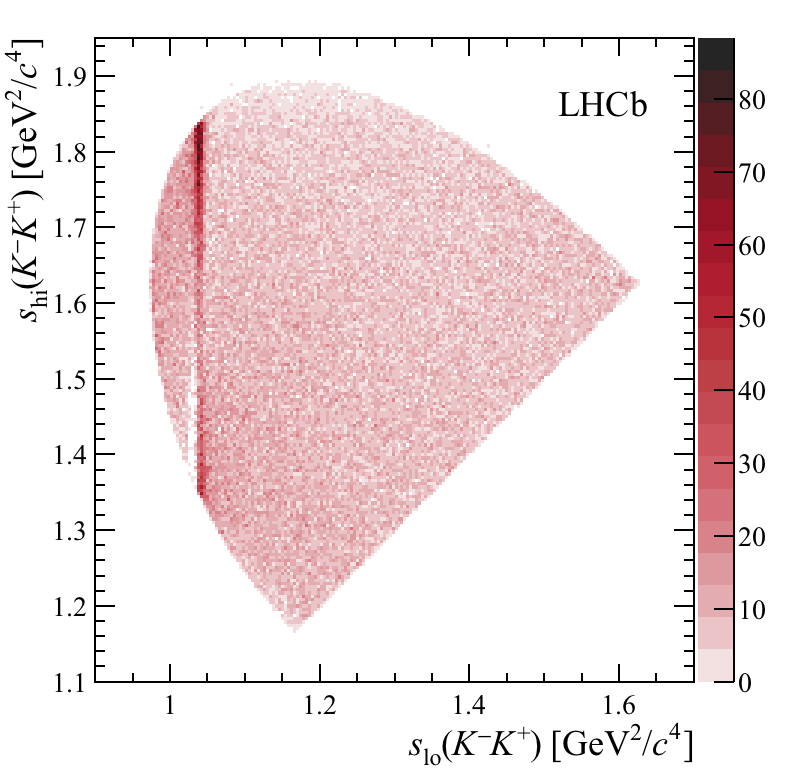}
\includegraphics*[width=.47\textwidth]{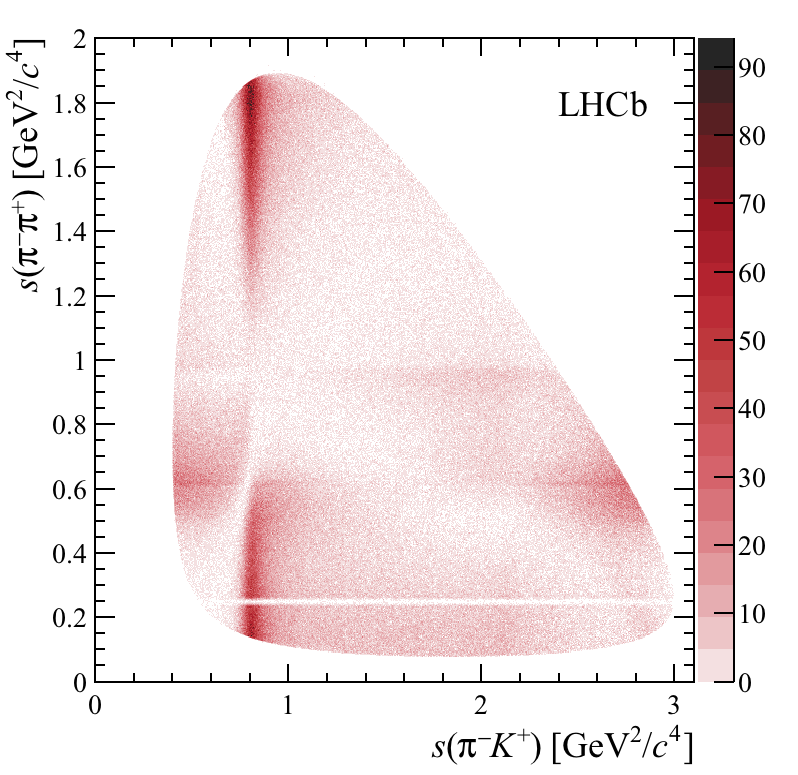}
\includegraphics*[width=.47\textwidth]{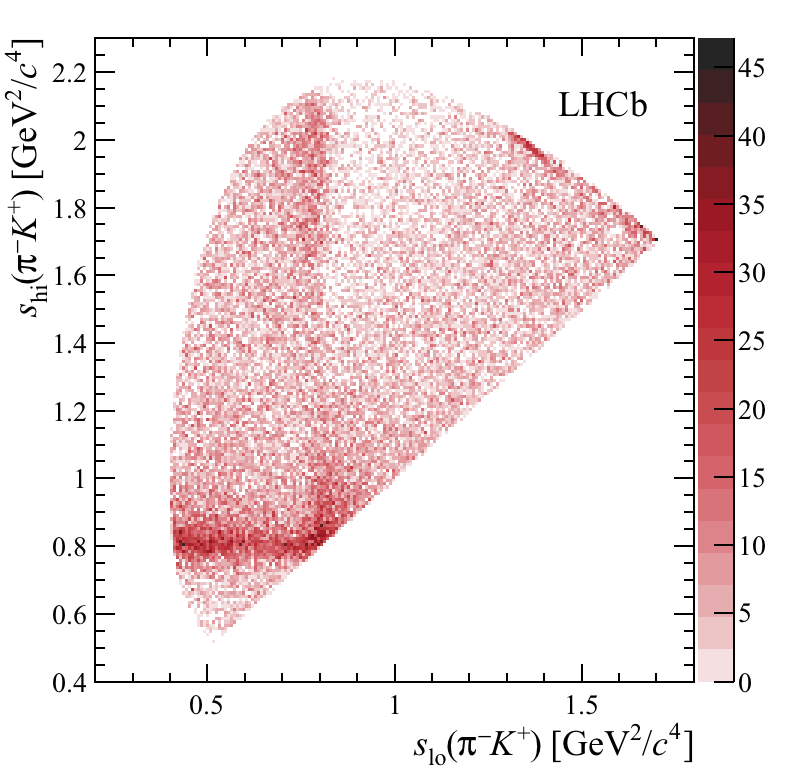}
\includegraphics*[width=.47\textwidth]{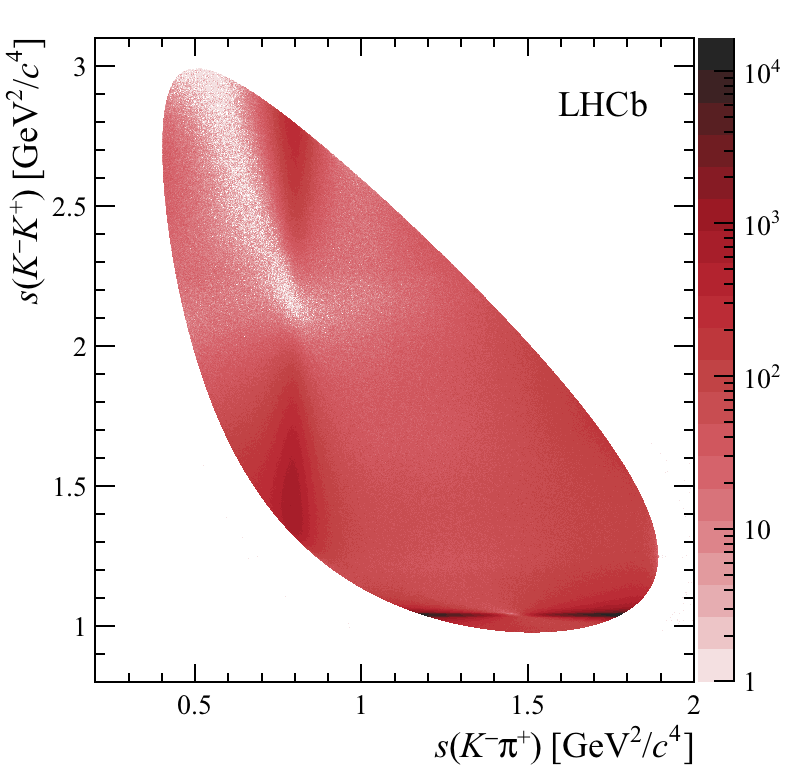}
\includegraphics*[width=.47\textwidth]{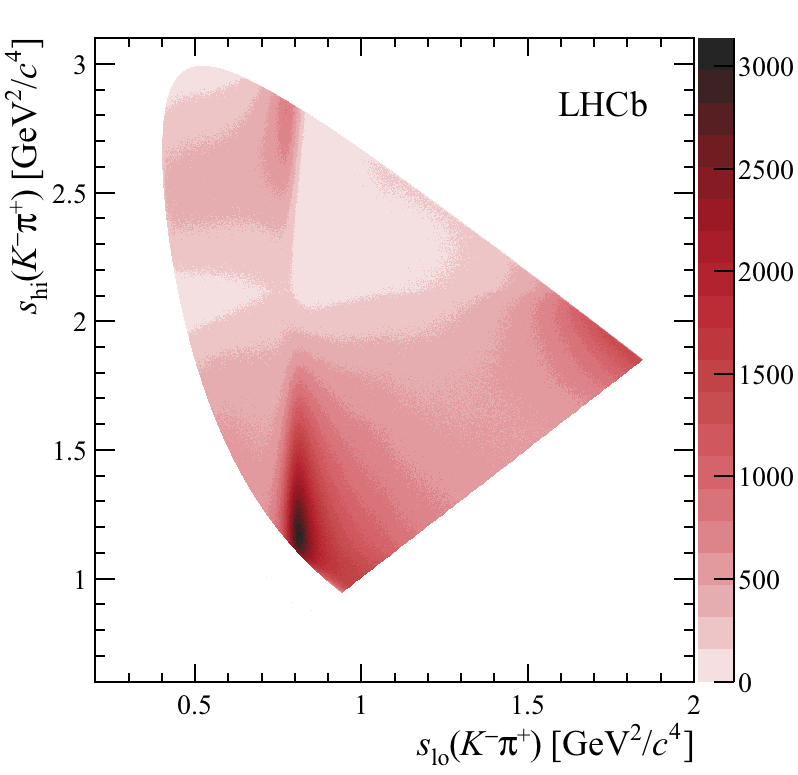}
\includegraphics*[width=.47\textwidth]{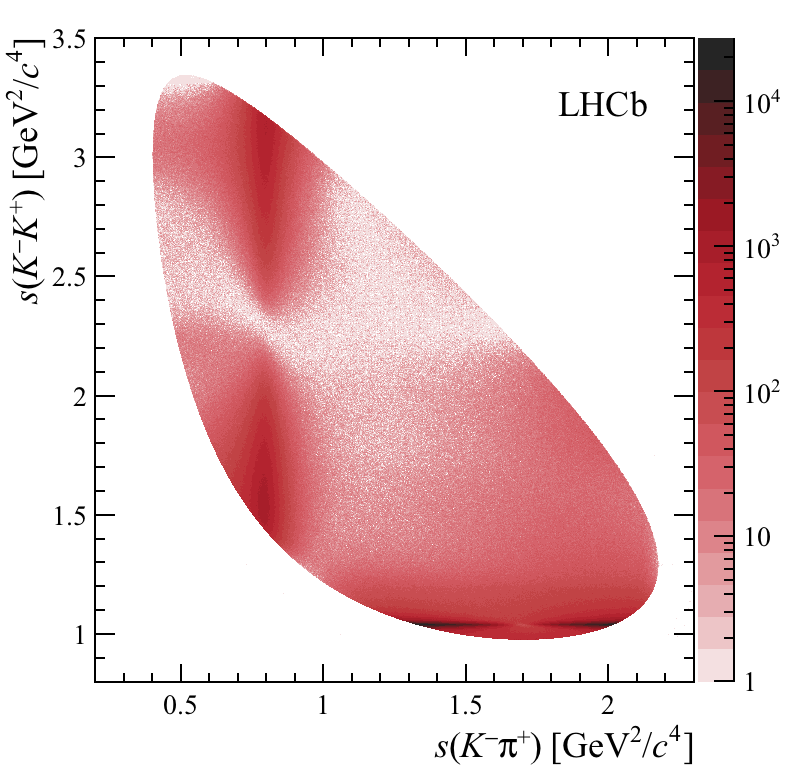}
\end{center}
    \vspace*{-.7cm}
\caption{\small Dalitz plots of the (top left) \dkkk, (top right) \dpipik, where the \KS veto can be seen, (middle left)  \dspikk, (middle right) \dkkpi, (bottom left) \dkpipi and (bottom right) \dskkpi  decays,  with  signal weights from \sPlot. A logarithmic scale is used for the \dkkpi and \dskkpi channels.}
  \label{fig:dps}
\end{figure}

The observed signal yields in bins of the DP, $N_{i}^{\rm obs}$ in Eq.~\ref{eq:nprod}, are determined using the \sPlot\ technique~\cite{Pivk:2004ty}. For each data subset, the signal and background \sWeights\ are obtained from the maximum-likelihood fit, and the former  are used to compute the number of signal candidates in each bin of the phase space for each data subset. No significant correlation between the $D^+_{(s)}$ candidate mass and  position in the DP is observed.  The DP with the total signal yields for all decays (merging \Dps and \Dms,  \MagDown and \MagUp subsets) are shown in Fig.~\ref{fig:dps}.

With the yields of observed candidates and the efficiencies obtained in bins of the DP for signal and normalisation modes, the total yields produced in the $pp$ collisions are evaluated using Eq.~\ref{eq:nprod}. These numbers are listed in Table~\ref{tab:prodYields}, separately for the \MagDown and \MagUp samples and can then be used for the determination of the different ratios of branching fractions.

\begin{table}[tb]
\caption{\small Produced yields for each decay mode  with statistical uncertainties, shown separately for \MagDown and \MagUp samples. The numbers given for the decay \dkpipi $^{(\dagger)}$ correspond to the sample with cuts optimised for the  \dkkk and \dkkpi and those for the decay \mbox{\dkpipi {\small $^{(\dagger\dagger)}$}} to the sample with cuts optimised for the \dpipik selection.}
 \vspace*{-0.5cm}
\begin{center}
\begin{tabular}{l r@{\,$\pm$\,}l r@{\,$\pm$\,}l }\hline
& \multicolumn{4}{c}{Produced yields $[\times 10^7]$} \\
      Decay         & \multicolumn{2}{c}{\MagDown}                        & \multicolumn{2}{c}{\MagUp} \\ \hline
 \dkkk              & 10.52&0.06& 10.54 &0.06 \\
 \dpipik            & 84.2 &0.4& 84.4  & 0.4 \\
 \dspikk            & 9.85  &0.15& 10.04  & 0.17 \\
 \dkkpi             & 1659  &12& 1651   & 13  \\
 \dkpipi{\small ($^{\dagger}$)}        &16103  & 40  & 16092  &40\\
 \dkpipi{\small ($ˆ{\dagger\dagger}$)} &16130  & 50 & 16101  &50\\
 \dskkpi                          & 4221  & 34 & 4150   &33\\\hline
 \end{tabular}
 \end{center}
 \label{tab:prodYields}
 \end{table}

\section{Systematic uncertainties}
\label{sec:systematics}

 Due to the similar decay topologies and selections applied to signal and normalisation channels within independent data subsets, many systematic uncertainties related to the final state factorise and cancel in the ratio of signal to normalisation yields and efficiencies.

The systematic uncertainties due to the limited size of the simulation samples are determined using pseudoexperiments. The number of generated events in each pseudoexperiment is obtained randomly in bins of the DP according to a Poisson distribution and corrected by the nominal  efficiency and correction maps due to PID, tracking and trigger. The uncertainties of the signal and normalisation efficiencies are taken as the Gaussian width of the resulting distributions of average  efficiencies. The mean value of these distributions are compatible with the nominal values. The resulting uncertainties of the ratios of efficiencies are given in Table~\ref{tab:Reff} and the corresponding relative systematic  uncertainties are given in Table~\ref{tab:sysMagUpDown}.

In order to estimate the uncertainties on the ratios of effective efficiencies arising from the limited size of the calibration samples used to determine the PID efficiency and the tracking and trigger corrections, 100 tables are generated, with efficiencies or corrections fluctuating according to Gaussian  functions centred at the nominal value and with width equal to their nominal uncertainties. For each generated table, the DP map of efficiencies are re-evaluated for signal and normalisation channels using the same procedure as the one used for the determination of the nominal efficiency maps. The distribution of the efficiency ratio is fitted with a Gaussian function, whose width is taken as systematic uncertainty.

 \begin{table}[t]
 \caption{\small Relative systematic uncertainties for the \MagDown and \MagUp results (in \%). The statistical uncertainties are also given for comparison.}
\vspace{-0.7cm}
 \begin{center} 
 \begin{tabular}{l  cccc }\hline
Source        &$\frac {\BF(\dkkk)} {\BF(\dkpipi)}$&  $\frac {\BF(\dpipik)} {\BF(\dkpipi)}$ & $  \frac {\BF(\dspikk)} {\BF(\dskkpi)}$ & $  \frac {\BF(\dkkpi)} {\BF(\dkpipi)}$  \\\hline
& & & & \\
&\multicolumn{4}{c}{\MagDown}\\\hline
 Size of simulation   & 0.34  &  0.47   & 1.0   &  0.75    \\
 PID          & 0.022 &  0.019  & 0.022  &  0.013        \\
 Tracking      & 0.22  &  0.069  & 0.079  &  0.11  \\
 Trigger corr.     & 0.011 & 0.0025  &  0.0050& 0.0057 \\
 Mat. description  & 0.53  &  $-$    & $-$    &  0.27  \\
 Fit Model    & 0.14  &  0.03   & 0.64   &  0.06    \\
 Sec. decays  & 0.18  & 0.25    & 0.31    & 0.11     \\
 DP Binning   & 0.09  &  0.05   & 0.30   &  0.13    \\ \hline
 Total syst.& 0.72  & 0.54    & 1.3    & 0.82         \\\hline
 Statistical   & 0.54  &  0.25   & 1.4    &  0.03 \\ \hline 
& & & & \\
&\multicolumn{4}{c}{\MagUp}\\\hline
 Size of simulation   &  0.32 &  0.52   & 1.2    &  0.81  \\
 PID          &  0.030&  0.020  & 0.023   &  0.021      \\
 Tracking      & 0.22  &  0.070  & 0.080  &  0.10       \\
 Trigger corr.     & 0.011  & 0.0024 &  0.0057& 0.0060  \\
 Mat. description  & 0.53  &  $-$    & $-$    &  0.27   \\
 Fit Model    &  0.13 &  0.07   & 0.54    &  0.06       \\
 Sec. decays  &  0.18 &  0.24   & 0.38    &  0.09       \\
 DP Binning      &  0.11 &  0.03   & 0.07    &  0.28     \\\hline 
 Total syst. & 0.71  & 0.58    & 1.3    & 0.91         \\\hline
Statistical   & 0.54  &  0.25   & 1.4     &  0.03 \\ \hline 
 \end{tabular}
\label{tab:sysMagUpDown}
 \end{center}
 \end{table}

 An additional systematic uncertainty is assigned to the ratios of tracking efficiencies when signal and normalisation decay modes have a different number of kaons and pions in the final state. The fractions of kaons and pions which cannot be reconstructed due to hadronic interactions that occur before the last tracking station are estimated using a simulated sample of \dkpipi decays. Assuming a 10\% uncertainty on the description of the detector material~\cite{LHCb-DP-2013-002}, per-track uncertainties on the efficiency of kaons and pions of (1.432$\pm$0.015)\% and (1.702$\pm$0.011)\%, respectively, are obtained. The residual uncertainties due to the different interactions of particles with opposite charge with the detector material are estimated to be negligible when compared to the uncertainty due to the limited size of the calibration samples, since the final branching-fraction ratios are averaged over particle charge and magnet polarity. This is the most important source of systematic uncertainty for the ${\BF(\dkkk)}/{\BF(\dkpipi)}$ measurement.

 The systematic uncertainty due to the fit model is estimated using an alternative parametrisation for the signal based on the sum of two CB functions with a common mean. 
 The observed yields obtained with this model are used to measure the branching-fraction ratios with the same procedure as for the nominal evaluation and the difference between the two results is assigned as systematic uncertainty.

The effect of residual charm contamination is studied. The stringent PID requirements are chosen to suppress the charm backgrounds to minimal levels, so that any remaining contribution  does not affect the signal yields,  either because the number of candidates is very low or because its shape is broad enough to be absorbed in the yield of the combinatorial background. This assumption is tested by explicitly estimating the residual contaminations and their shapes from data and simulation, and including them in the mass fits. No significant effects are found for any of the signal modes.
The impact of the mass vetoes used to reject the \Lc contamination is studied by further enlarging the  mass-veto window by 5\mevcc for all channels. No  significant deviation is observed in any of the final results and therefore no systematic uncertainty is assigned.

The effect of a potential contamination from decays of \Dps from \bquark-hadron decays, which could be  different for signal and normalisation samples, is investigated by tightening the requirement on \chisqip\ to two alternative values and measuring the ratios of branching fractions. The largest deviation from the nominal value is assigned as systematic uncertainty. 

The systematic uncertainty due to the choice of DP binning scheme is evaluated as the deviation of the ratio of produced yields obtained with alternative binning schemes from that obtained with the nominal binning schemes. These binning schemes are defined by changing the nominal number of bins by $\pm$1, $\pm$2 and $\pm$4 units in each DP axis. 

The systematic uncertainties due to the different sources considered in this analysis are summarised in Table~\ref{tab:sysMagUpDown}, separately for the \MagDown and \MagUp results.  Except for the ${\BF(\dkkk)}/ {\BF(\dkpipi)}$ measurement, the most important source of systematic uncertainty is the limited size of the simulation samples. However, the only result with total uncertainty dominated by this contribution is the branching-fraction ratio of the CS decay \dkkpi.

The ratios of branching fractions obtained with data taken with the two magnet polarities are shown  in Table~\ref{tab:BRsperPol}, with statistical and systematic uncertainties. For each decay mode the two results are compatible and no additional systematic uncertainty is assigned to the effect of detector asymmetry. 

\begin{table}[t]
\caption{\small Ratios of branching fractions, shown separately for \MagDown and \MagUp samples. The first uncertainty is statistical and the second is systematic.}
\vspace{-0.7cm}
\begin{center}
\begin{tabular}{l r@{\,$\pm$\,}l@{\,$\pm$\,}l}\hline
 Channel &  \multicolumn{3}{c}{\MagDown ($\times 10^{-3}$)  }      \\ \hline
 \footnotesize ${\BF(\dkkk)}/ {\BF(\dkpipi)}$   & 0.653& 0.004& 0.005 \\
 \footnotesize ${\BF(\dpipik)}/ {\BF(\dkpipi)}$ & 5.220 & 0.013 & 0.028  \\
 \footnotesize ${\BF(\dspikk)}/ {\BF(\dskkpi)}$ & 2.333 & 0.033 & 0.030  \\
 \footnotesize ${\BF(\dkkpi)}/ {\BF(\dkpipi)}$  & 103.00& 0.03  & 0.85   \\ \hline
  Channel &  \multicolumn{3}{c}{\MagUp ($\times 10^{-3}$)  }      \\ \hline
  \footnotesize ${\BF(\dkkk)}/ {\BF(\dkpipi)}$   & 0.655& 0.004& 0.005  \\
  \footnotesize ${\BF(\dpipik)}/ {\BF(\dkpipi)}$ & 5.244 & 0.013 & 0.030  \\
  \footnotesize ${\BF(\dspikk)}/ {\BF(\dskkpi)}$ & 2.419 & 0.035 & 0.032  \\
  \footnotesize ${\BF(\dkkpi)}/ {\BF(\dkpipi)}$ & 102.59& 0.03  & 0.93   \\ \hline
  \end{tabular}
  \end{center}
  \label{tab:BRsperPol}
  \end{table}

\section{Results}
\label{sec:results}

Final ratios of branching fractions are obtained by combining the two measurements  shown in Table~\ref{tab:BRsperPol}, accounting for 100\% correlation~\cite{Nisius:2014wua} between the systematic uncertainties due to the material description in the simulation, fit model, contamination from secondary decays and DP binning. For the doubly Cabibbo-suppressed channels, the results are

\begin{align}
\frac {\BF(\dkkk)} {\BF(\dkpipi)}& =  (6.541 \pm 0.025  \pm 0.042) \times 10^{-4}, \nonumber \\
 \frac {\BF(\dpipik)} {\BF(\dkpipi)}& = (5.231 \pm 0.009  \pm 0.023) \times 10^{-3}, \nonumber\\
  \frac {\BF(\dspikk)} {\BF(\dskkpi)} &= (2.372 \pm 0.024  \pm 0.025) \times 10^{-3}, \nonumber
\end{align}
\noindent where the first uncertainty is statistical and the second systematic. 
These values are consistent with the current world averages, being compatible at the 1.4$\sigma$, 2.4$\sigma$ and 0.2$\sigma$ levels, respectively. 
 
 In addition, the result for the Cabibbo-suppressed mode \dkkpi is

$$ \frac {\BF(\dkkpi)} {\BF(\dkpipi)} = (10.282 \pm 0.002 \pm 0.068)\times 10^{-2}, $$
\noindent where again the first uncertainties are statistical and the second systematic.
It is in agreement with the world average~\cite{PDG2018} at the 1.6$\sigma$ level, improving the precision by a factor 2.6.

 The ratios of branching fractions are combined with the world-average
 values~\cite{PDG2018} of the branching fractions of the CF decays \dkpipi ($8.98\pm 0.28$)\% and \dskkpi ($5.45 \pm 0.17$)\% to compute the branching fractions of the DCS modes
 \begin{align}
 {\BF(\dkkk)} & =  (5.87 \pm 0.02  \pm 0.04 \pm 0.18) \times 10^{-5}, \nonumber \\
 {\BF(\dpipik)} & = (4.70 \pm 0.01  \pm 0.02 \pm 0.15) \times 10^{-4}, \nonumber\\
  {\BF(\dspikk)} &= (1.293 \pm 0.013  \pm 0.014 \pm 0.040) \times 10^{-4}, \nonumber
\end{align}
and  of the CS mode

 $${\BF(\dkkpi)}  = (9.233 \pm 0.002 \pm 0.061 \pm 0.288)\times 10^{-3},$$
 
 \noindent where the uncertainties are statistical, systematic and due to the uncertainty of the normalisation channel, respectively. Altogether, these represent the best measurements up to date.

\section*{Acknowledgements}
%
% These Acknowledgements valid from 14-Aug-2018
%
\noindent We express our gratitude to our colleagues in the CERN
accelerator departments for the excellent performance of the LHC. We
thank the technical and administrative staff at the LHCb
institutes.
We acknowledge support from CERN and from the national agencies:
CAPES, CNPq, FAPERJ and FINEP (Brazil); 
MOST and NSFC (China); 
CNRS/IN2P3 (France); 
BMBF, DFG and MPG (Germany); 
INFN (Italy); 
NWO (Netherlands); 
MNiSW and NCN (Poland); 
MEN/IFA (Romania); 
MSHE (Russia); 
MinECo (Spain); 
SNSF and SER (Switzerland); 
NASU (Ukraine); 
STFC (United Kingdom); 
NSF (USA).
We acknowledge the computing resources that are provided by CERN, IN2P3
(France), KIT and DESY (Germany), INFN (Italy), SURF (Netherlands),
PIC (Spain), GridPP (United Kingdom), RRCKI and Yandex
LLC (Russia), CSCS (Switzerland), IFIN-HH (Romania), CBPF (Brazil),
PL-GRID (Poland) and OSC (USA).
We are indebted to the communities behind the multiple open-source
software packages on which we depend.
Individual groups or members have received support from
AvH Foundation (Germany);
EPLANET, Marie Sk\l{}odowska-Curie Actions and ERC (European Union);
ANR, Labex P2IO and OCEVU, and R\'{e}gion Auvergne-Rh\^{o}ne-Alpes (France);
Key Research Program of Frontier Sciences of CAS, CAS PIFI, and the Thousand Talents Program (China);
RFBR, RSF and Yandex LLC (Russia);
GVA, XuntaGal and GENCAT (Spain);
the Royal Society
and the Leverhulme Trust (United Kingdom);
Laboratory Directed Research and Development program of LANL (USA).

\addcontentsline{toc}{section}{References}
\setboolean{inbibliography}{true}
\bibliographystyle{LHCb}
\bibliography{main,LHCb-PAPER,LHCb-CONF,LHCb-DP,LHCb-TDR,standard}

\newpage
                                                                                                                      
\centerline{\large\bf LHCb collaboration}
\begin{flushleft}
\small
R.~Aaij$^{27}$,
C.~Abell{\'a}n~Beteta$^{44}$,
B.~Adeva$^{41}$,
M.~Adinolfi$^{48}$,
C.A.~Aidala$^{76}$,
Z.~Ajaltouni$^{5}$,
S.~Akar$^{59}$,
P.~Albicocco$^{18}$,
J.~Albrecht$^{10}$,
F.~Alessio$^{42}$
% ...

{\footnotesize \it
% ...
$ ^{5}$Clermont Universit{\'e}, Universit{\'e} Blaise Pascal, CNRS/IN2P3, LPC, Clermont-Ferrand, France\\
% ...
$ ^{10}$Fakult{\"a}t Physik, Technische Universit{\"a}t Dortmund, Dortmund, Germany\\
% ...
$ ^{18}$INFN Laboratori Nazionali di Frascati, Frascati, Italy\\
% ...
$ ^{27}$Nikhef National Institute for Subatomic Physics, Amsterdam, Netherlands\\
% ...
$ ^{41}$Instituto Galego de F{\'\i}sica de Altas Enerx{\'\i}as (IGFAE), Universidade de Santiago de Compostela, Santiago de Compostela, Spain\\
$ ^{42}$European Organization for Nuclear Research (CERN), Geneva, Switzerland\\
% ...
$ ^{44}$Physik-Institut, Universit{\"a}t Z{\"u}rich, Z{\"u}rich, Switzerland\\
% ...
$ ^{48}$H.H. Wills Physics Laboratory, University of Bristol, Bristol, United Kingdom\\
% ...
$ ^{59}$University of Cincinnati, Cincinnati, OH, United States\\
% ...
$ ^{61}$Syracuse University, Syracuse, NY, United States\\
% ...
$ ^{76}$University of Michigan, Ann Arbor, United States, associated to $^{61}$\\
% ...
}
\end{flushleft}

\end{document}